\documentclass{aa}  

\usepackage{graphicx}
\usepackage{txfonts}
\usepackage[colorlinks=true, linkcolor=blue, citecolor=blue, urlcolor=blue]{hyperref}
\begin{document}

\title{The spatially-resolved effect of mergers on the stellar mass assembly of MaNGA galaxies}

   \author{Eirini Angeloudi\inst{1, 2}
          \and
          Marc Huertas-Company\inst{1, 2, 3, 4}
          \and 
          Jesús Falcón-Barroso\inst{1, 2}
          \and
          Laurence Perreault-Levasseur\inst{5, 6, 7} 
          \and 
          Alexandre Adam\inst{5, 6}
          \and
          Alina Boecker\inst{8}
          }

   \institute{Instituto de Astrof\'isica de Canarias, C. V\'ia L\'actea, 1, E-38205 La Laguna, Tenerife, Spain
         \and
             Universidad de la Laguna, dept. Astrof\'isica, E-38206 La Laguna, Tenerife, Spain
        \and
           Universit\'e Paris-Cit\'e, LERMA - Observatoire de Paris, PSL,          Paris, France
        \and
            SCIPP, University of California, Santa Cruz, CA 95064, USA
        \and
            D\'epartement de Physique, Universit\'e de Montr\'eal, Succ. Centre-Ville, Montr\'eal, Qu\'ebec, H3C 3J7, Canada
        \and
            Mila—Quebec Artificial Intelligence Institute, Montreal, Qu\'ebec, Canada
        \and
            Center for Computational Astrophysics, Flatiron Institute, NY, USA
        \and
            Department of Astrophysics, University of Vienna, T\"urkenschanzstrasse 17, 1180 Vienna, Austria
    }

   \date{Received; accepted}

 
\abstract
{Understanding the origin of stars within a galaxy -- whether formed in-situ or accreted from other galaxies (ex-situ) -- is key to constraining its evolution. Spatially resolving these components provides crucial insights into a galaxy’s mass assembly history.}
{We aim to predict the spatial distribution of ex-situ stellar mass fraction in MaNGA galaxies, and to identify distinct assembly histories based on the radial gradients of these predictions in the central regions.}
{We employ a diffusion model trained on mock MaNGA analogs (MaNGIA), derived from the TNG50 cosmological simulation. The model learns to predict the posterior distribution of resolved ex-situ stellar mass fraction maps, conditioned on stellar mass density, velocity, and velocity dispersion gradient maps. After validating the model on an unseen test set from MaNGIA, we apply it to MaNGA galaxies to infer the spatially-resolved distribution of their ex-situ stellar mass fractions - i.e. the fraction of stellar mass in each spaxel originating from mergers.}
{We identify four broad categories of ex-situ mass distributions: (1) flat gradient, in-situ dominated; (2) flat gradient, ex-situ dominated; (3) positive gradient; and (4) negative gradient. The vast majority of MaNGA galaxies fall in the first category -- flat gradients with low ex-situ fractions -- confirming that in-situ star formation is the main assembly driver for low- to intermediate-mass galaxies. At high stellar masses $(>10^{11} M_\odot)$, the ex-situ maps are more diverse, highlighting the key role of mergers in building the most massive systems. Ex-situ mass distributions correlate with morphology, star-formation activity, stellar kinematics, and environment, indicating that accretion history is a primary factor shaping massive galaxies. Finally, by tracing their assembly histories in TNG50, we link each class to distinct merger scenarios, ranging from secular evolution to merger-dominated growth.}
{The central gradients of the ex-situ stellar mass fraction encode meaningful information about the assembly history of galaxies. Our results highlight the power of combining cosmological simulations with machine learning to infer the unseen components of galaxies from observable properties.}


   \maketitle
%
\section{Introduction} \label{sec:intro}

Despite the vastness of the Universe, galaxies tend to cluster in relatively small regions \citep{1980lssu.book.....P}, and gravity can bring them close enough to merge. Galaxy mergers play a fundamental role in galaxy evolution, serving as a primary mechanism for increasing a galaxy’s stellar mass and size \citep{2015ARA&A..53...51S}. Particularly for massive galaxies, this process is part of a two-phase formation scenario \citep{2010ApJ...725.2312O}, in which galaxies first grow rapidly through in-situ star formation at high redshift (z > 2), followed by a prolonged phase of ex-situ stellar mass assembly through the accretion of stars formed in other galaxies, primarily via merger events. These mergers occur in various forms: a galaxy may absorb smaller satellite galaxies, gradually increasing its stellar mass \citep{2009ApJ...699L.178N}, or galaxies of comparable size may collide and merge, through a more violent, high-impact event \citep{2012A&A...548A...7L, 2013MNRAS.428..999P, 2019ApJ...876..110D}.

In addition to increasing stellar mass, mergers can have a variety of pronounced effects in the evolution of a galaxy. They can trigger morphological transformations \citep{2009ApJS..182..216K}, induce inflows of gas that rejuvenate star formation \citep{2011MNRAS.412..591P}, and fuel AGN activity \citep{2008AJ....135.1877E}. Yet, quantifying how much of a galaxy’s stellar mass is built through mergers remains challenging, as disentangling the origin of stars to in-situ and ex-situ stellar components is not directly observable.

A series of simulation studies have helped establish global trends in ex-situ stellar mass buildup across cosmic time. \cite{2010ApJ...725.2312O}, using cosmological zoom-in simulations, found that the ex-situ stellar mass fraction strongly correlates with galaxy mass, with more massive galaxies typically having assembled a larger portion of their stars through mergers. This trend has since been confirmed in both semi-analytic models \citep{2013MNRAS.434.3348C, 2015MNRAS.451.2703C} and large-volume cosmological simulations \citep{2016MNRAS.458.2371R, 2019MNRAS.490.3234N, 2020MNRAS.497...81D, 2022ApJ...935...37R}. Across these studies, a consistent picture emerges: ex-situ stellar mass fractions increase with galaxy stellar mass, conforming to a hierarchical scenario of galaxy evolution. Additional trends with galaxy characteristics emerge, e.g. accretion is more significant in the stellar mass assembly of systems with early-type morphologies and slow rotation \citep{2017MNRAS.467.3083R, 2019MNRAS.487.5416T}. These global trends reflect the cumulative role of mergers in shaping galaxies but do not capture the full spatial complexity of accreted material.

More recently, a number of studies have taken advantage of the detailed information provided by cosmological simulations to track how accreted mass is deposited spatially within galaxies. In general, these works find that in-situ stars dominate the inner regions, while accreted stars are preferentially deposited at larger radii, leading to an average positive gradient in the ex-situ stellar mass fraction radial profile \citep{2016MNRAS.458.2371R, 2017MNRAS.464.2882A, 2019MNRAS.487..318K, 2020MNRAS.497...81D, 2023MNRAS.519.5202B}. Cosmological simulations have also revealed that the spatial distribution of accreted material strongly depends on the mass ratios and timing of merger events. More specifically, minor and very minor mergers typically fail to reach the central regions, instead contributing to the buildup of the stellar halo \citep{2016MNRAS.458.2371R}. This is consistent with observational constraints that identify minor mergers as the primary driver of the late-time size growth in early-type galaxies \citep{2009ApJ...699L.178N}. In contrast, more massive satellites, which experience stronger dynamical friction, can sink deeper into the host's potential. Temporal signatures are also imprinted in the radial distribution: stars from earlier mergers are deposited at smaller radii, while those from more recent events populate the outskirts \citep{2017MNRAS.464.2882A}. This transition radius between in-situ and ex-situ dominance varies systematically with stellar mass, morphology, and assembly history \citep{2016MNRAS.458.2371R, 2022ApJ...935...37R}.

While these simulation results are insightful, similar observational constraints remain challenging and often rely on support from theoretical models. In the case of the Milky Way, understanding its accreted stellar halo is a central goal of Galactic Archaeology - a field that reconstructs the Galaxy’s merger history through detailed chemo-dynamical analysis of halo stars \citep{2020ARA&A..58..205H}. Within the Local Group, several studies have combined deep photometric mapping of stellar structures of galaxies in the Milky Way stellar mass regime with N-body simulations of satellite accretion to estimate their accreted mass content \citep{2017MNRAS.466.1491H, 2022ApJ...930...69S}. Beyond the Local Group, more indirect methods have been applied. Approaches using surface brightness profile decompositions and stellar population gradients - mainly from large integral field spectroscopy surveys - have attempted to constrain ex-situ content \citep{2017A&A...603A..38S, 2018MNRAS.475.3348H, 2019ApJ...880..111O, 2020MNRAS.491..823B, 2021MNRAS.507.3089D, 2023MNRAS.520.5651C}. These studies often rely on simplified models or comparisons to simulations and typically find that accreted stars dominate at large radii in massive early-type galaxies, similar to results found in large-volume simulations. \looseness-2 

Recently, the dynamically hot inner halo has been proposed as an observational proxy for the ex-situ stellar mass. Using the TNG50 simulation, \cite{2022A&A...660A..20Z} demonstrated that this component traces the cumulative accreted fraction reasonably well. This framework was then applied to estimate the ex-situ content in two early-type galaxies in the Fornax cluster \citep{2022A&A...664A.115Z}. However, this approach requires computationally expensive orbital decomposition methods and thus its application remains limited so far in terms of sample size. Beyond structural tracers, observational estimates of merger rates provide another constraint on the contribution of mergers to galaxy growth \citep{2015A&A...576A..53L, 2016ApJ...830...89M, 2017MNRAS.470.3507M, 2018MNRAS.475.1549M, 2025MNRAS.tmp..779P}. \citet{2022ApJ...940..168C} used deep near-infrared imaging from the REFINE survey to directly measure galaxy pair fractions and merger rates out to $z \sim 3$, quantifying the stellar mass growth driven by both major and minor mergers. This analysis was recently extended by \cite{2025MNRAS.tmp..606D} using JWST data, showing that mergers contribute to roughly half of the stellar mass buildup at high redshift, reinforcing their dominant role in early galaxy assembly. These results offer an independent measure of how mergers shape galaxies across mass and time.

While the majority of these works focus on retrieving global trends, a vast pool of information is encoded in the spatial distribution of the accreted stellar mass. Recent studies using the IllustrisTNG \citep{2021A&A...647A..95P} and Magneticum \citep{2022ApJ...935...37R} simulations have shown that the spatial imprint of ex-situ stars can be linked to the detailed merger history of galaxies. In particular, these works defined subgroups of galaxies based on the shape of their radial ex-situ fraction profiles - such as steeply rising, flat, or declining - and connected these profile types to differences in merger histories, including the number, timing, and mass ratios of past mergers. These findings offer valuable insights into the assembly of galaxies, but translating them into real, observational constraints remains non-trivial.

Machine learning offers a promising avenue to bridge this gap \citep{2024A&A...687A..24M, 2021arXiv210205182K, 2025MNRAS.538L..31F, 2024A&A...687A..45P}. By combining the predictive capabilities of neural networks with the detailed ground truth available in simulations, one can infer physical quantities that are otherwise inaccessible in observational data. Galaxy mergers have been widely studied with the power of machine learning, from detecting tidal features \citep{2019MNRAS.483.2968W}, to identifying merging phases \citep{2021MNRAS.504..372B}, estimating merger rates \citep{2020ApJ...895..115F}, and probing mergers at higher redshifts \citep{2020A&C....3200390C}. In our recent work \citep{2023MNRAS.519.2199E, 2023MNRAS.523.5408A}, we focused on a more challenging aspect: retrieving information on the past stellar mass assembly of galaxies from their observable features through a "merger archaeology" approach. We showed that neural networks can retrieve key merger-related parameters from simulated galaxies using realistic observational inputs. Specifically, we demonstrated that spatially-resolved maps of stellar mass and kinematics from integral field unit (IFU) data can be used to predict ex-situ stellar mass fractions with high accuracy across two different cosmological simulations. Other efforts using photometric data have reported similar results \citep{2025A&A...695A.177C}, further supporting the potential of this approach.

Building on that foundation, our latest work trained a machine learning model on MaNGIA (mock MaNGA analogs) \citep{2023A&A...673A..23S} and applied it to real MaNGA galaxies to estimate their global ex-situ stellar mass fractions \citep{2024NatAs...8.1310A}. This revealed, for the first time in a statistically significant observational sample, trends between ex-situ content and stellar mass, morphology, star formation, and environment. In the present study, we expand this framework by training a more advanced machine learning model to predict two-dimensional, spatially-resolved ex-situ stellar mass fraction maps. These maps allow for enhanced insight into the spatial structure of the accreted component and its relation to each galaxy’s merger history. Our methodology provides a robust pathway toward understanding the role of mergers in galaxy evolution, particularly in disentangling the contributions of in-situ and ex-situ processes.

The paper is structured as follows. In Section 2, we introduce both the observational dataset (MaNGA) and the synthetic dataset (MaNGIA), mocked like MaNGA and used to train a machine learning model that infers 2D spatially-resolved maps of the ex-situ stellar mass fraction. Section 3 details the creation of ground-truth maps from simulations and the architecture of the diffusion model. In Section 4, we validate the method on mock test data. In Section 5, we proceed by applying the diffusion model to the MaNGA galaxy sample, generating ex-situ fraction maps for over 10,000 galaxies, which are analyzed via their radial profiles and classified into four groups based on central gradients. In Section 6, we link these groups to distinct merger histories using TNG50 and propose four distinct evolutionary scenarios. We summarize our conclusions in Section 7.

\section{Datasets} \label{sec:datasets}

\subsection{MaNGA}
MaNGA \citep{2015ApJ...798....7B} (Mapping Nearby Galaxies at Apache Point Observatory) is the integral field spectroscopic component of SDSS-IV, designed to spatially map the kinematic and spectroscopic properties of over 10,000 nearby galaxies. The sample spans stellar masses from $10^{9} M_\odot$ to $10^{12} M_\odot$ and redshifts between $0.01 < z < 0.15$ (median $z \sim 0.03$), with spectral coverage from 3600–10300 Å at a resolution of R~$\sim$~2000. Each galaxy is observed using a fiber bundle with diameters ranging from 12'' (19 fibers) to 32'' (127 fibers), depending on the target size. The full sample is divided into three sub-samples: the Primary sample (covering out to 1.5$R_e$, $50\%$ of targets), the Secondary sample (up to 2.5$R_e$, $\sim 33\%$ of the targets), and the Color-Enhanced sample ($\sim 17\%$), which supplements underpopulated regions in the NUV–$i$ vs $M_i$ color–magnitude plane \citep{2017AJ....154...86W}. Stellar mass estimates are taken from the NSA (NASA-Sloan Atlas) catalog, based on K-corrected elliptical Petrosian fluxes. To ensure consistency with the simulations, we correct these masses to reflect the Planck 2015 cosmology \citep{2016A&A...594A..13P}, adjusting from the NSA’s $h = 1$ to $h = 0.6774$ using:
$M^\star_{corr} = M^\star - 5\textrm{log}_{10}(h)/2.5$. We also utilize various pre-measured properties from the MaNGA survey:

\begin{itemize}
    \item Morphological Classification: Galaxies are categorized into three types (elliptical, S0, spiral) using the MaNGA PyMorph morphological catalogue \citep{2022MNRAS.509.4024D}. We identify 2460 elliptical, 923 S0 and 5311 spiral galaxies in the MaNGA sample.
    \item Star-Formation Activity:
    We classify galaxies as star-forming or quenched using the integrated star-formation rate (SFR) from the Ha emission line, as derived from the PIPE3D analysis \citep{2016RMxAA..52..171S}. The separation is made on the SFR vs. stellar mass plane, resulting in 5724 star-forming and 4519 quenched galaxies. 
    \item Rotation and Angular Momentum:
    Galaxies are classified as slow or fast rotators using the $\lambda_{Re}$ angular momentum proxy and ellipticity  $\epsilon$ , based on the criterion proposed by \cite{2016ARA&A..54..597C} and data from \cite{2018MNRAS.477.4711G}. We identify 964 slow rotators and 9279 fast rotators.
    \item  Halo mass:
     We use halo mass estimates from the self-calibrating halo-based group catalog \citep{2021ApJ...923..154T}. We identify 3038 galaxies in haloes with $M_{\rm halo} > 10^{13} M_\odot$, 2616 galaxies in the range $10^{12} M_\odot< M_{\rm halo} < 10^{13} M_\odot$, and 4589 galaxies in haloes with $M_{\rm halo} < 10^{12} M_\odot$.
     \item Environment: We use the environment classification provided by the MaNGA GEMA-VAC  \citep{ArgudoFernandezInPrep}. Using the classification scheme provided, we identify 2905 galaxies residing in clusters, 3793 in filaments, 989 in sheets and 140 in voids.
\end{itemize}

\subsection{MaNGIA}

In this work, we use MaNGIA \citep{2023A&A...673A..23S} as our training set. MaNGIA is a forward-modeled dataset based on the TNG50 simulation \citep{2019MNRAS.490.3196P, 2019MNRAS.490.3234N}, which represents the highest-resolution run within the IllustrisTNG suite. Galaxies are selected from TNG50 to match the stellar mass, size, and redshift distribution of the MaNGA survey and are forward-modelled to resemble realistic MaNGA-like integral field datacubes. This process includes synthetic spectral generation, emulation of the MaNGA fibre bundle geometry, and incorporation of observational effects such as seeing and instrumental noise. The final datacubes are processed with the PIPE3D pipeline \citep{2022ApJS..262...36S, 2022NewA...9701895L}, enabling a consistent analysis of stellar populations and kinematics. Since PIPE3D is also used in the official MaNGA data releases, it allows for direct comparison between mock and observed galaxies. The MaNGIA dataset accurately reproduces key observed trends in galaxy properties, including metallicity gradients, stellar age distributions, and velocity fields, making it well-suited for data-driven studies. In addition to the publicly released sample, we expand our training dataset with an extended set of $\sim$1600 simulated galaxies generated through the same methodology, not yet incorporated in the public dataset. \looseness-2

\section{Methods} \label{sec:methods}

\subsection{Ex-situ 2D Maps Creation}
The goal of this work is to train a machine learning model capable of predicting the 2D distribution of ex-situ stellar mass fraction of galaxies from their spatially resolved observables, specifically stellar mass and kinematics maps. We adopt a supervised learning approach, where the target ground-truth data are derived from large-volume cosmological simulations. 

Since MaNGIA is based on galaxies from the TNG50 run of the IllustrisTNG simulation suite, we use the simulation's particle-level information to define the ex-situ stellar mass. Following the definition in \cite{2016MNRAS.458.2371R}, stellar particles are classified as in-situ if they formed within the main progenitor branch of the galaxy they currently belong to, and as ex-situ if they formed outside of it. This classification has already been applied to all stellar particles in TNG50 and is available through the stellar assembly catalogs \citep{2016MNRAS.458.2371R}, making MaNGIA a unique dataset for exploring the spatial imprint of galaxy accretion histories.

For each MaNGIA galaxy, we extract the stellar particles within the field of view used for the original forward modelling and apply the same projection. Using a 2D grid, we spatially bin the stellar particles, summing the stellar masses of stars within each bin (separating per origin) to create resolved maps of the ex-situ and in-situ stellar mass distributions. The ex-situ stellar mass fraction map is then obtained by dividing the ex-situ mass map by the sum of the ex-situ and in-situ mass maps in each bin. Then, these resolved ex-situ stellar mass fraction maps serve as the training targets for our machine learning model. 

\subsection{Diffusion models}

\begin{figure}
    \centering
    \includegraphics[width=\linewidth]{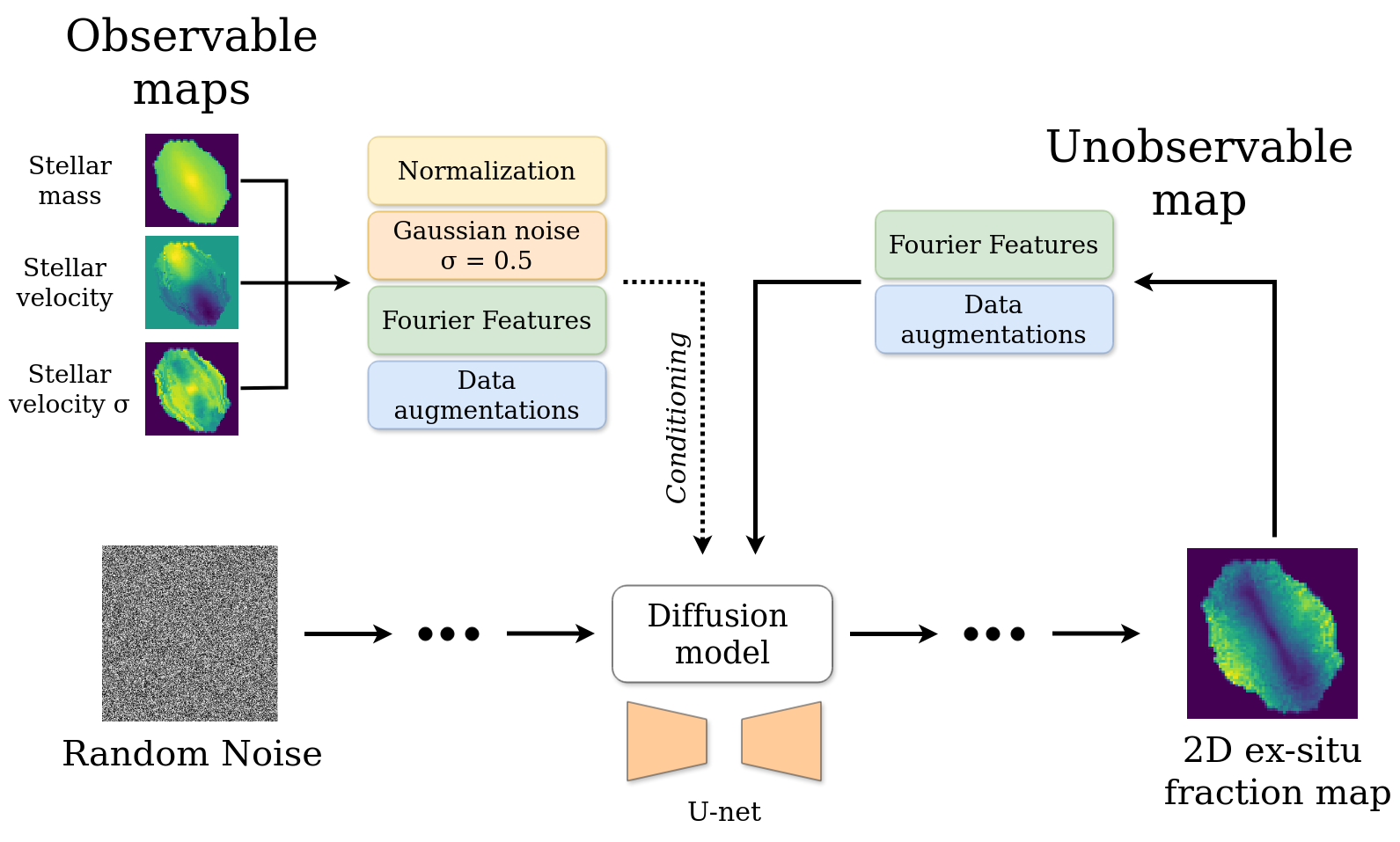}
    \caption{Schematic of the neural network architecture used in this work to predict 2D spatially resolved maps of the ex-situ stellar mass fraction from observable 2D maps of stellar mass density, velocity, and velocity dispersion. A diffusion model is trained to learn the underlying distribution of ex-situ maps using the MaNGIA training set, conditioned on the corresponding observable maps (after a series of preprocessing steps, for details see Section \ref{preprocessing}).}
    \label{fig:chart}
\end{figure}

In our previous work \citep{2024NatAs...8.1310A}, we demonstrated the use of simulation-based inference to predict the global ex-situ stellar mass fraction of MaNGA galaxies from IFU-like data. That study employed a simple convolutional neural network (CNN), enhanced with self-supervised techniques to improve generalization. In the present work, we build upon that approach by extending the prediction task to two-dimensional, spatially-resolved ex-situ stellar mass fraction maps. This added complexity requires a more sophisticated machine learning architecture.\looseness-2

Diffusion models have recently emerged as a powerful class of generative models, achieving state-of-the-art performance in high-quality image/video generation, as well as text-to-image translation, by learning to reverse a gradual noising process applied to data \citep{2020arXiv201113456S}. Recently, they have also seen increasing use in astrophysics — particularly in inverse problems and in modeling complex, high-dimensional data distributions. Notable applications include generating realistic galaxy images \citep{2022MNRAS.511.1808S}, reconstructing maps of matter distribution \citep{2023A&A...672A..51R}, modeling gravitational lensing systems \citep{2022mlps.workE...1A, 2022arXiv221104365K}, reconstructing high-quality observational images across various domains \citep{2023arXiv230509121W, 2023arXiv230411751F, 2024ApJ...975..201F, 2024A&A...683A.105D}, and accurately characterizing non-Gaussian, anisotropic noise in space-based observations \citep{2023ApJ...949L..41L, 2025AJ....169..254A}. 

At a high level, diffusion models learn the underlying data distribution by performing two complementary processes: a forward process, where noise is gradually added to a data point (e.g. an image) until it becomes indistinguishable from pure noise and a reverse process, where noise is iteratively removed to generate new samples from the learned distribution. Crucially, the denoising process is not random — it is guided by the score function, which is defined as the gradient of the logarithm of the data probability density $p(x)$ with respect to the input data $x$:

\begin{equation}
    \centering
    \text{score} = \nabla_x \log p(x)
\end{equation}

This quantity, called the score, points in the direction where the data distribution increases most rapidly in data space. In other words, it tells us how to move $x$ in order to increase its likelihood under the data distribution $p(x)$. A neural network is typically trained to learn the score distribution using Score Matching \citep{JMLR:v6:hyvarinen05a, 10.1162/NECO_a_00142}. In our case, we use denoising score matching predicting the objective from \citet{2019arXiv190705600S}. The noising and denoising processes are described by underlying stochastic differential equations (SDEs). For a detailed mathematical description, we refer the reader to the original work by \citet{2020arXiv201113456S, 2021arXiv210109258S}.\looseness-2

In addition to unconditional generation, diffusion models can be adapted for conditional generation, where the model is guided to produce samples consistent with specific input information. In our case, the model is conditioned on 2D observable IFU-like maps of stellar mass, velocity, and velocity dispersion, enabling us to generate ex-situ stellar mass fraction maps that are consistent with observed galaxy properties. In the current work, we use the score-based models from the repository \texttt{score\_models} \footnote{The code is available at \url{https://github.com/AlexandreAdam/score_models}.}. A schematic of our architecture can be found in Fig. \ref{fig:chart}.

\subsection{Preprocessing} \label{preprocessing}

Before training the diffusion model on the MaNGIA dataset, we apply some preprocessing steps both to the observable maps (conditions) and the ex-situ stellar mass maps. First, we normalize the condition maps by standardizing each IFU map individually. This normalization removes any global information, leaving only the local gradients of the maps. By stripping away mass dependencies, the conditional maps represent the variations in the data, which the diffusion model learns to relate to the spatially-resolved ex-situ stellar mass fractions. We note here that the same normalization has been applied in our previous works \citep{2023MNRAS.523.5408A, 2024NatAs...8.1310A}, when the same kind of observable maps proved robust when the model was calibrated to predict the global ex-situ stellar mass fraction across two cosmological simulations.
 
Additionally, we incorporate Fourier features (FF) into the diffusion model \citep{2021arXiv210700630K, 2024arXiv240800839B} to enhance both the conditions and labels. Specifically, we apply Fourier transformations to the IFU-like images (conditions) and the 2D ex-situ fraction maps (labels), generating high-frequency sine and cosine components corresponding to both the first and second-order frequencies. These components are concatenated and input into the model. By including this kind of Fourier features, the model can capture more spatial variations and high-frequency details in both the observed maps and target labels, improving its ability to handle complex, high-dimensional data in the diffusion process.

Finally, we apply specific cuts to the MaNGIA dataset based on total stellar mass and global ex-situ stellar mass fraction. We include only synthetic galaxies with stellar mass $>10^{9} M_\odot$ and $f_{ex-situ, *} > 0.05$. This is to ensure some balancing in our dataset in terms of the ex-situ stellar mass fraction distributions. Without this cut, the dataset would be biased towards low, flat ex-situ stellar mass fraction maps (zero everywhere), complicating the training process. We end up with $\sim 6300$ galaxies that we split into training set (90\%) and test set (10\%). We note here that the synthetic galaxies in the MaNGIA dataset may be result of the same object from the simulation after applying different projections. We ensure that all projections of the same subhalo are in the same split to avoid data leaks between the training and the validation step. \looseness-2

\begin{figure*}
    \centering
    \includegraphics[width=\linewidth]{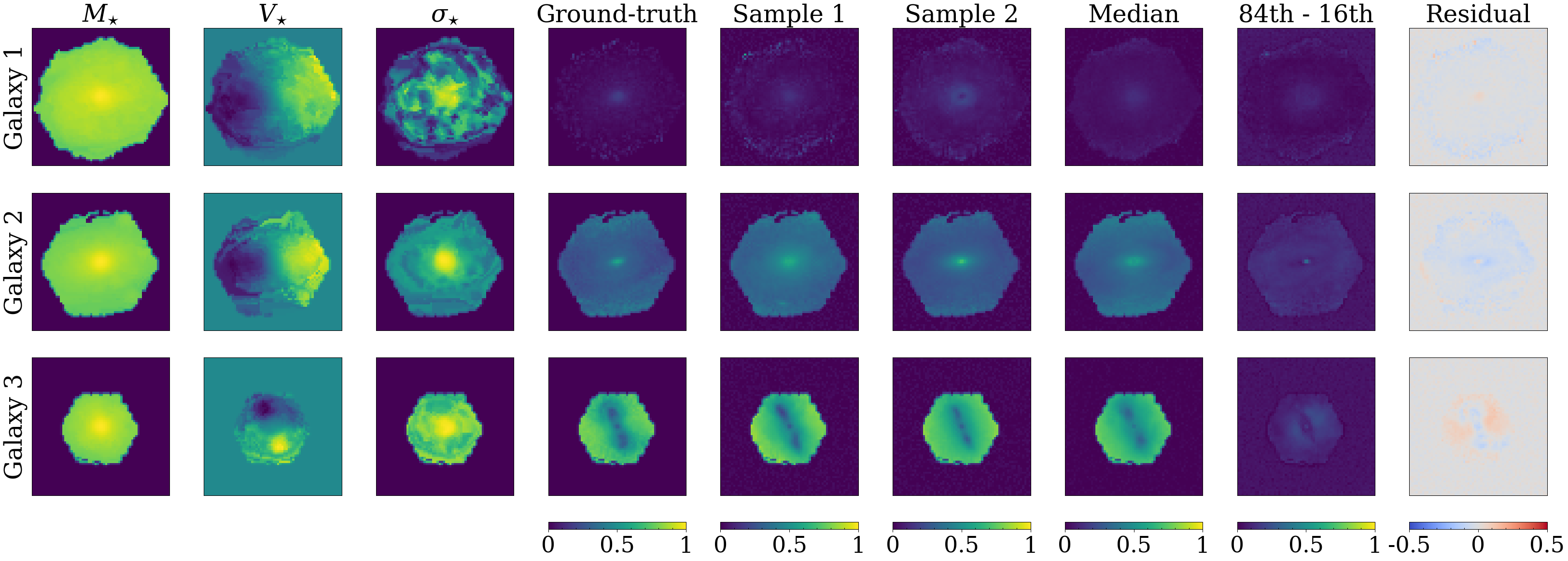}

    \caption{Reconstruction of the ground-truth from the diffusion model for 3 mock galaxies from the MaNGIA test set. The first three columns display the input observable maps provided to the diffusion model: stellar mass, stellar velocity, and velocity dispersion. These maps have been forward-modeled through the MaNGIA pipeline and subsequently normalized to retain only gradient information. The fourth column shows the ground-truth 2D ex-situ stellar mass fraction map from the TNG50 cosmological simulation. The fifth and sixth columns present two individual samples from the diffusion model. The seventh column displays the median prediction from 100 samples. Finally, the last two columns show the model uncertainty (computed as the 84th–16th percentile range) and the residual between the median prediction and the ground-truth map, respectively.  }
    \label{fig:reconstruction_examples}
\end{figure*}

\begin{figure*}
    \centering
    \includegraphics[width=\linewidth]{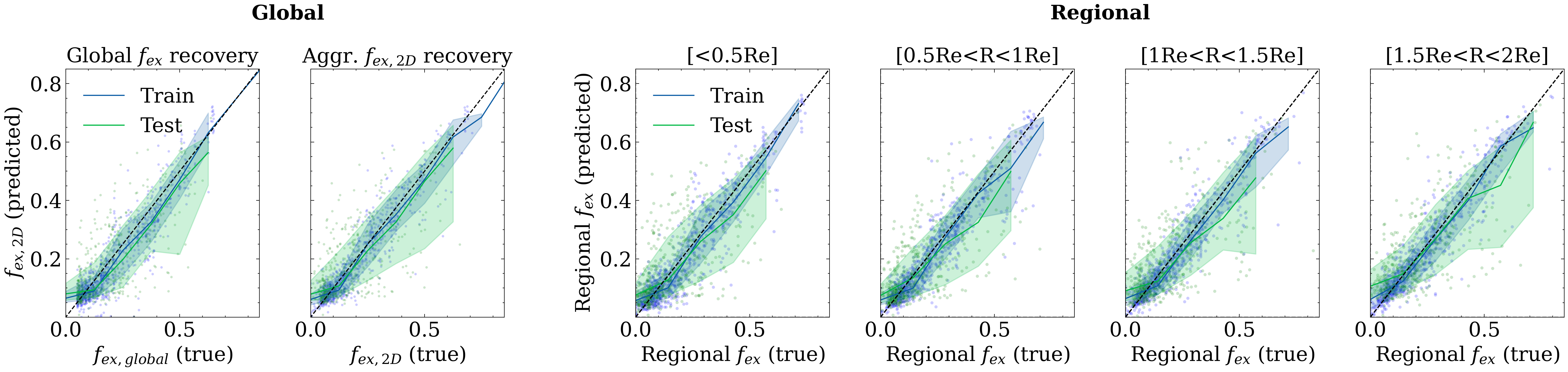}
    \caption{Metrics of accuracy of the diffusion model trained on the MaNGIA dataset. The recovery of the global ex-situ stellar mass fraction from 2D predicted map of the model vs. the true global value of the ex-situ stellar mass fraction for the train and test set (left). The aggregated ex-situ stellar mass fraction from the predicted 2D map vs. the aggregated value from the ground-truth 2D map for the train and test set (middle). The recovery of the regional trends in the predicted map vs. the ground-truth map for the train and test set integrated at different annuli (right). The green solid line denotes the median recovery, and the shaded region covers 68 percent of the test set, representing the prediction scatter. For reference, we also include the same comparison for 1,000 training set galaxies in blue, underlining that the model is not overfitting.}
    \label{fig:model_metrics}
\end{figure*}

This dataset is relatively small for the task at hand, however a possible extension is not viable due to the time-consuming mocking procedure. Nevertheless, we find that the model is still able to effectively capture the gradients in the ex-situ stellar mass distributions for the majority of galaxies, as will be discussed in the following section. To improve generalization and mitigate overfitting, we apply several data augmentation techniques to the conditions. These include horizontal and vertical flips, random rotations, and random resized crops with varying scales and aspect ratios. Additionally, we introduce Gaussian noise with a standard deviation of $\sigma = 0.5$ to the conditions. After adding the noise, the conditions are standardized again to maintain consistency in the input data.

\section{Validation on the MaNGIA test set} \label{sec:validation}

We train the diffusion model using the designated training set and evaluate its performance on a separate test set reserved for validation. For each test galaxy, we generate 100 samples of the 2D ex-situ stellar mass fraction map, conditioned on the three observable input maps: stellar mass density, velocity, and velocity dispersion, all normalized to contain only spatial gradients. We adopt the median of the 100 samples as the final prediction, and use the 84th–16th percentile range to quantify uncertainty.

In Figure \ref{fig:reconstruction_examples}, we show example reconstructions for three typical galaxies from the test set.  Overall, we find that the model is able to reconstruct the ground truth ex-situ stellar mass fraction maps with high accuracy. Despite being conditioned only on a limited set of physically informative inputs, it successfully captures complex spatial features across a wide range of ex-situ distributions. The model recovers both low and high ex-situ fractions in different galactic regions with high accuracy, and residuals remain low throughout.

Apart from inspecting individual 2D map reconstructions, we also assess the model’s overall performance across the test set by evaluating its ability to recover both global and spatially-resolved ex-situ stellar mass information. For the global evaluation, we compare the aggregated ex-situ stellar mass fraction derived from the predicted 2D maps to both the true global value and the value obtained from the ground-truth 2D maps. To compute the aggregated ex-situ fraction from the predicted maps, we first recover the ex-situ stellar mass at each pixel by multiplying the predicted fraction map with the corresponding stellar mass map. We then sum the total ex-situ mass across all pixels and divide by the total stellar mass.

In the first two columns of Figure \ref{fig:model_metrics}, we show for each galaxy in the test set, the comparison between the aggregated ex-situ stellar mass fraction from the predicted 2D map and the true global value (first panel). The second panel presents a complementary comparison between the predicted and true 2D aggregated ex-situ fractions, showing consistent agreement and confirming the model’s ability to capture global trends, even for unseen galaxies.

One key advantage of producing 2D maps rather than predicting only global quantities is the ability to probe the radial distribution of ex-situ stellar mass. To test the model’s ability to recover such spatial variations, we evaluate the aggregated ex-situ stellar mass fraction within several radial apertures, scaled by the galaxy’s effective radius. Each panel in the right hand-side of Figure \ref{fig:model_metrics} shows the comparison of predicted versus true ex-situ fractions in different radial bins. The calculation follows the same procedure as in the global case but restricted to pixels within each aperture. We can see that the diffusion model is able to recover not only the global information for the mock MaNGIA galaxies from the test set but also the spatial variations in the 2D ex-situ stellar mass fraction distributions. Notably, the model shows increasing accuracy toward the outer regions, where the ex-situ component typically dominates and displays more prominent gradients.

We conclude that the trained model is able to recover both the global as well as the regional ex-situ stellar mass fraction distributions, using only observable information from stellar mass and kinematic IFU maps.

\section{Application to MaNGA Galaxies} \label{sec:results}

\subsection{Comparison with 1D predictions}

Having demonstrated the effectiveness of the diffusion model in recovering the ex-situ stellar mass fraction both globally and in a spatially resolved manner from just three observable maps, we now apply our trained and validated model to real galaxy data from the MaNGA survey. The input conditions required to sample the diffusion model—stellar mass, velocity, and velocity dispersion maps—are available for the full sample of $\sim10,000$ MaNGA galaxies in the same format as the synthetic MaNGIA observables, as they were originally mocked for this purpose. Consequently, no additional preprocessing is necessary beyond the existing pipeline. After normalizing the MaNGA maps to retain only the spatial gradients, we supply them directly to the diffusion model. For each MaNGA galaxy, we generate 10 samples from the model, producing - for the first time - predictions of the 2D ex-situ stellar mass fraction distribution across this large observational dataset.

\begin{figure}
    \centering
    \includegraphics[width=\linewidth]{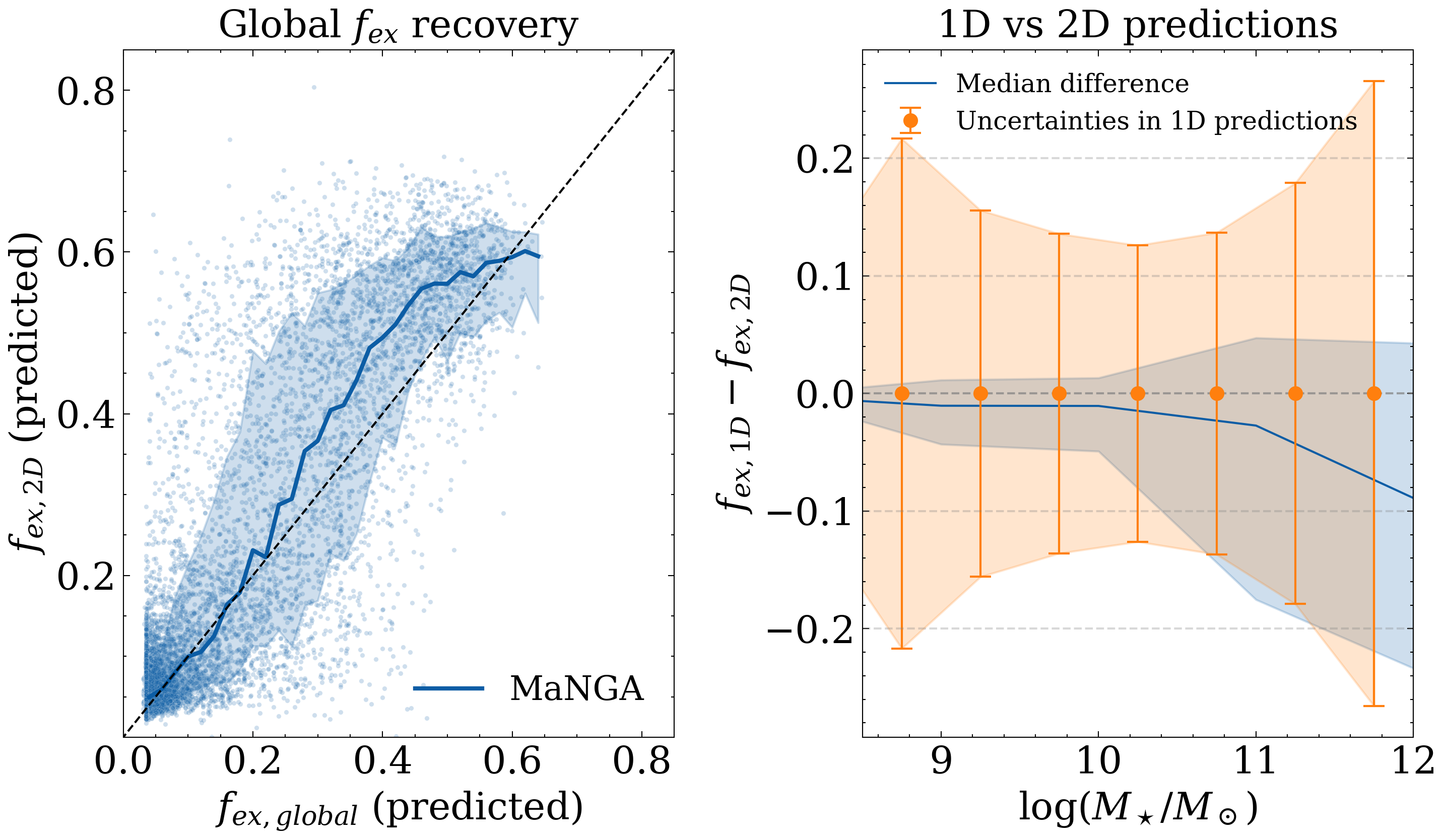}
    \caption{Comparison of the integrated 2D predictions with the 1D predictions from \cite{2024NatAs...8.1310A}. (a) The recovery of the global ex-situ stellar mass fraction from 2D predicted map of the model vs. the predicted global value of the ex-situ stellar mass fraction for the MaNGA dataset from previous work. Each point represents a galaxy, and the solid blue line traces the median of the distribution. (b) The median difference of the two predictions as a function of stellar mass. The orange error bars display the median uncertainty per stellar mass bin from the 1D predictions. While the residual between predictions shows a slight increase towards higher stellar masses, it is still covered by the produced uncertainties. The shaded regions enclose 68\% of all data.}
    \label{fig:manga_1D_2D}
\end{figure}

As a first verification of our predictions, we compare the aggregated ex-situ stellar mass fractions derived from the 2D maps predicted by the diffusion model to the global ex-situ fractions previously estimated for MaNGA galaxies using a CNN-based model \citep{2024NatAs...8.1310A}. In that earlier work, we used the same set of observables—stellar mass, velocity, and velocity dispersion maps—to directly predict the integrated ex-situ stellar mass fraction. In Figure \ref{fig:manga_1D_2D}, we show this comparison for the full MaNGA sample of 10,000 galaxies. We find that the relation lies close to the one-to-one line, with the diffusion model generally predicting slightly higher ex-situ fractions than the CNN. 

This mild offset is expected, given the differences in model architectures and the fact that the diffusion model was trained on a narrower stellar mass range. Moreover, the diffusion model captures higher-order spatial structure in the data, which the CNN is not designed to model, potentially contributing to the observed discrepancies. In the right panel of Figure~\ref{fig:manga_1D_2D}, we additionally show the difference between the two predictions as a function of stellar mass. The solid blue line reveals a trend toward larger deviations at the high-mass end, where the diffusion model tends to predict higher values of ex-situ fraction. However, this difference remains within the uncertainties produced from the 1D CNN-based estimates. The main trends identified in our previous work \citep{2024NatAs...8.1310A} remain consistent with the current results, as illustrated in Fig.~\ref{fig:recreation} in the Appendix~\ref{appendix_recreate}. Overall, this comparison provides a reassuring consistency between the two approaches and supports the validity of our diffusion model’s predictions.

\subsection{Radial profiles of ex-situ stellar mass fraction}

\begin{figure}
    \includegraphics[width=\linewidth]{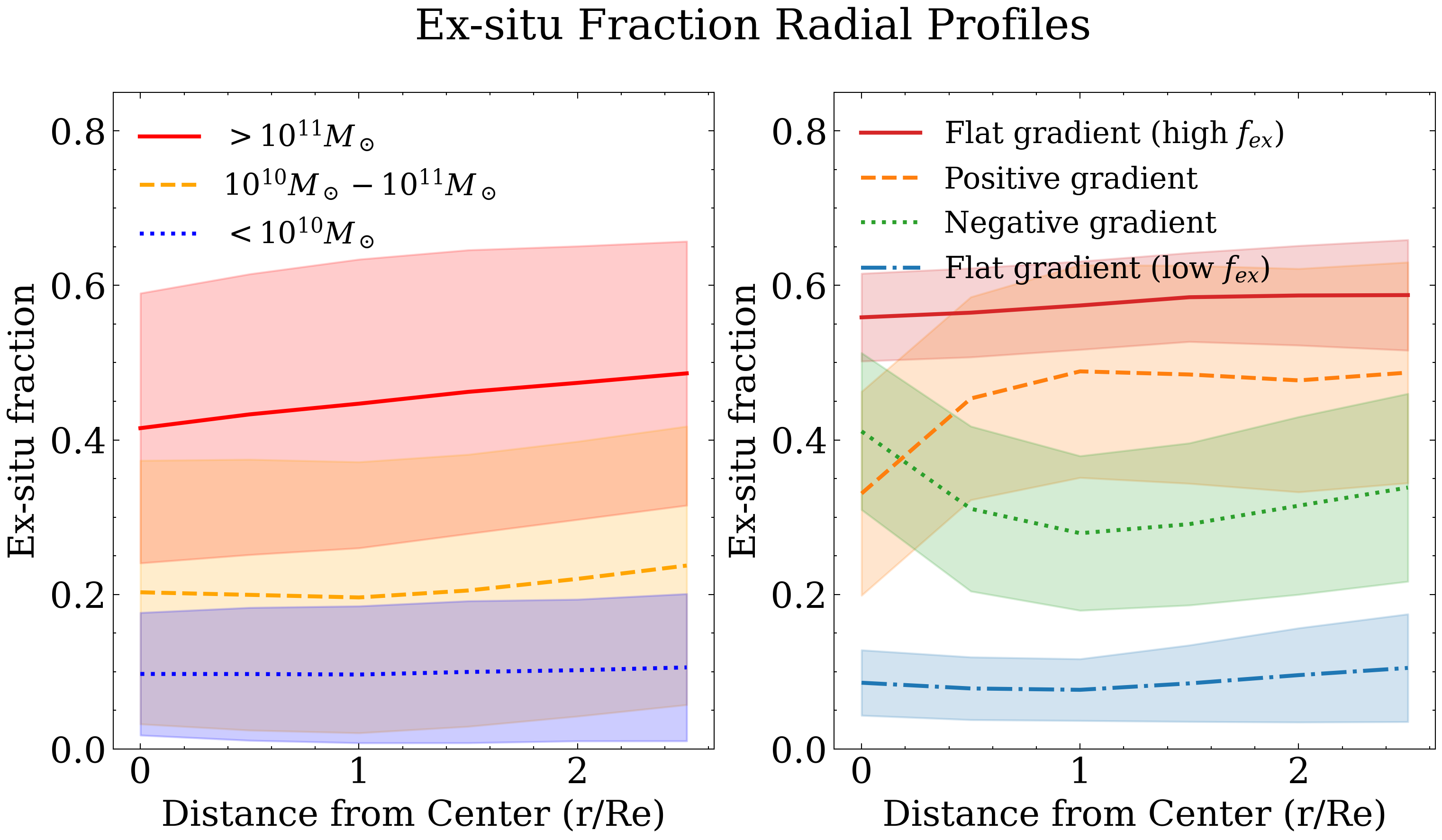} 
    \caption{The stacked radial profiles of the ex-situ stellar mass fraction for 10,000 MaNGA galaxies as predicted from the diffusion model. (a) The median of the stacked radial profiles separated in 3 stellar mass bins. (b) The median of the stacked radial profiles in each group, following the classification based on the central ex-situ gradient. The shaded regions enclose 68\% of all data.}
    \label{fig:radial_profiles}
\end{figure}

\begin{figure*}
    \centering
    \includegraphics[width=\linewidth]{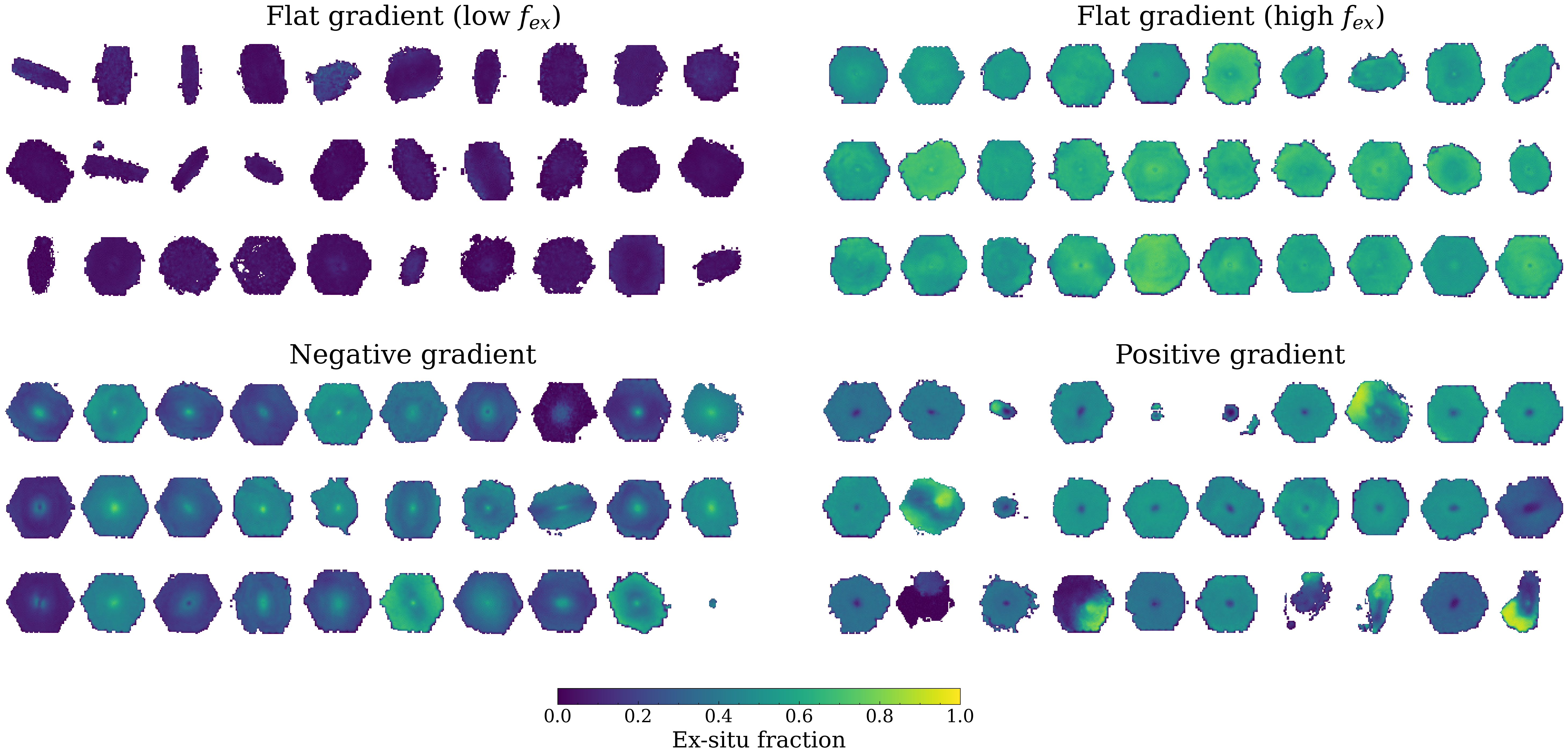}
    \caption{2D ex-situ stellar mass fraction maps for 30 randomly selected galaxies in each group, based on the classification developed in this work. Each map shows the median of 10 samples predicted by the diffusion model, trained on the MaNGIA mock dataset and conditioned on the stellar mass density and kinematic maps of the corresponding MaNGA galaxy. The different groups display different spatial distributions in their ex-situ stellar mass fraction, following our classification scheme.}
    \label{fig:slope_groups}
\end{figure*}

We continue our analysis by examining the spatial information encoded in the 2D ex-situ stellar mass fraction maps of the MaNGA galaxies. For each galaxy, we compute the radial profile of the ex-situ stellar mass fraction by calculating the fraction at different radii from the galaxy center, normalized by its effective radius ($R_\mathrm{e}$). We then stack the radial profiles of all galaxies within three stellar mass bins, and present the results in the left panel of Figure~\ref{fig:radial_profiles}.

We find that, overall, the radial profiles of the ex-situ stellar mass fraction are relatively flat across all stellar mass bins, and follow the expected trend of increasing global ex-situ fraction with stellar mass. More specifically, low-mass galaxies ($M_\star < 10^{10} M_\odot$) exhibit flat radial profiles with ex-situ stellar mass fractions around 0.1 across all radii. Intermediate-mass galaxies ($10^{10} M_\odot < M_\star < 10^{11} M_\odot$) also display flat profiles, but at higher values. High-mass galaxies ($M_\star > 10^{11} M_\odot$) show significantly higher ex-situ fractions, with central values around 0.4 that increase toward the outskirts, reaching $\sim 0.5$ at $2.5 R_\mathrm{e}$.

These trends are consistent with theoretical predictions from cosmological simulations \citep{2016MNRAS.458.2371R, 2020MNRAS.497...81D} and align with results from our previous work \citep{2024NatAs...8.1310A}, which established a strong correlation between stellar mass and the ex-situ stellar mass fraction. The slight increase in ex-situ fraction with radius also agrees with prior literature \citep[e.g.][]{2010ApJ...725.2312O, 2016MNRAS.458.2371R, 2021MNRAS.507.3089D, 2019ApJ...880..111O}, although previous studies often report steeper gradients and higher values in the outer regions. We emphasize, though, that our analysis primarily probes the central regions of galaxies, in contrast to most previous studies that extend to larger radii. Additionally, the stacking procedure used here may smooth out intrinsic variations, potentially removing strong radial gradients that exist in individual galaxies. \looseness-2

\subsection{Classes based on ex-situ fraction central gradient}

\begin{figure*}
    \centering
    \includegraphics[width=\linewidth]{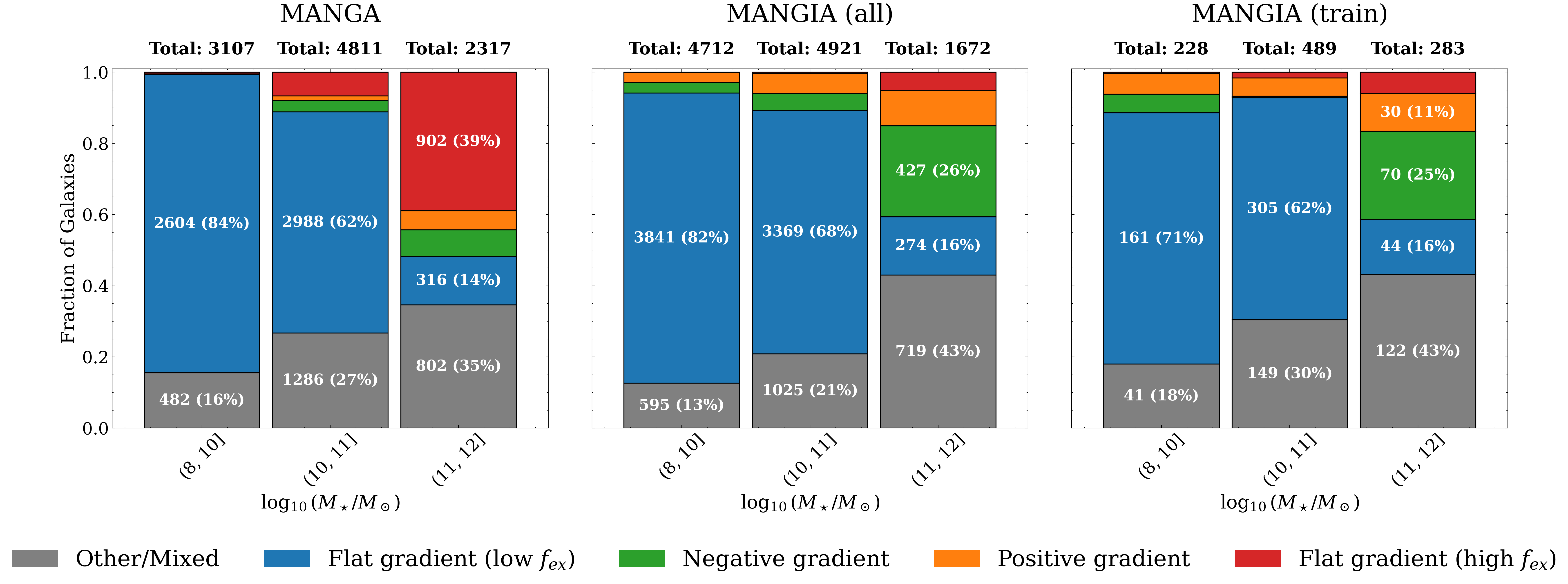}
    \caption{The abundances of the 5 different groups classified based on the ex-situ stellar mass fraction gradient in their inner parts for MaNGA, MaNGIA and the subest of MaNGIA used for training in 3 stellar mass bins. The percentages of each group in the respective bin are shown in white bold text and ommited if they are less that 10\%. The abundances of the groups are different between the observational data and the synthetic data that the diffusion model was trained on.}
    \label{fig:mass_fractions_groups}
\end{figure*}

\begin{figure}
    \centering
    \includegraphics[width=\linewidth]{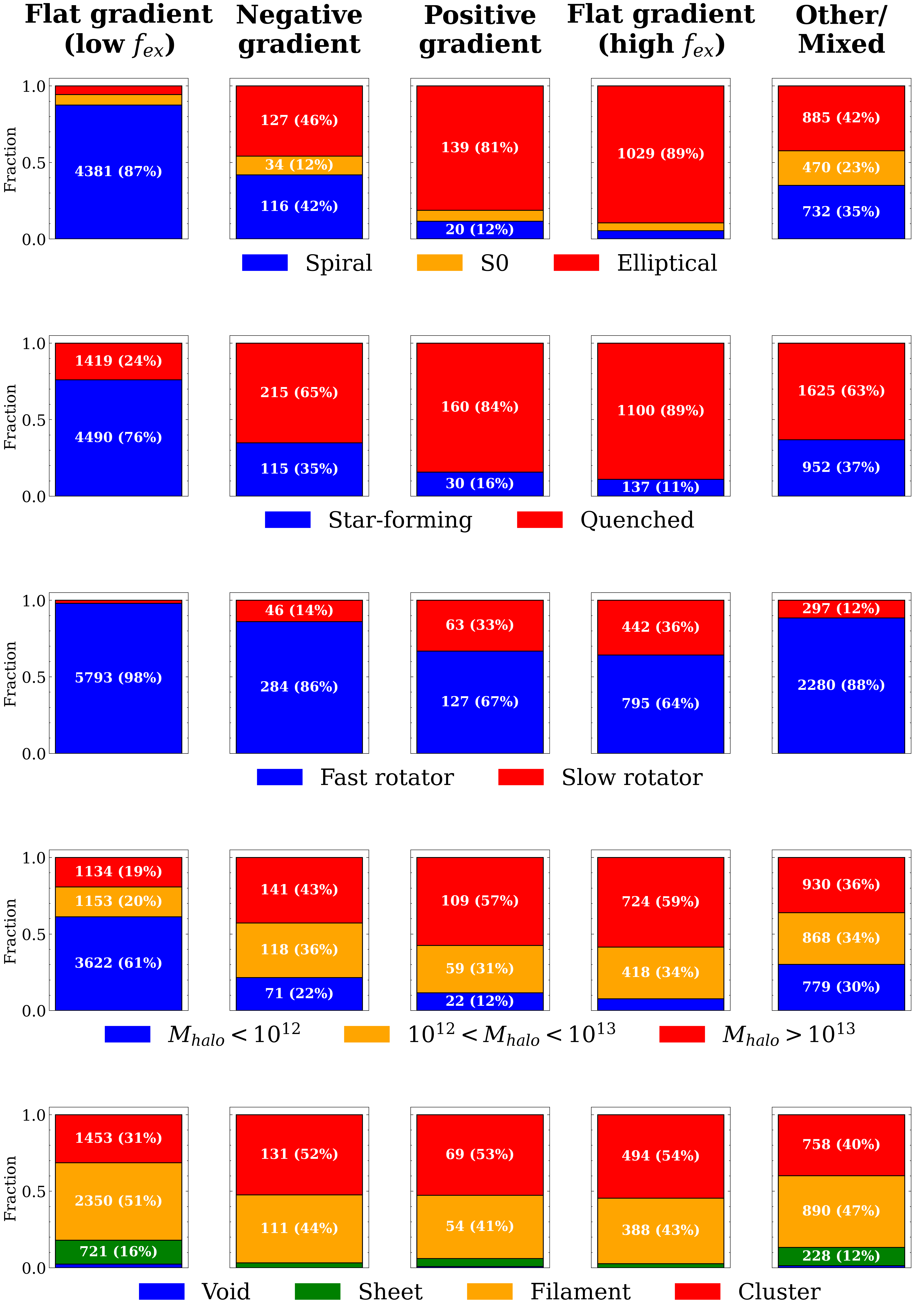}
    \caption{The abundances of different galaxy properties across the 5 groups of MaNGA galaxies classified based on the ex-situ stellar mass fraction gradient in their inner parts. The 5 classes show clear differences in morphology, star formation, rotation, and environment.}
    \label{fig:slope_group_trends}
\end{figure}

To better capture diversity and isolate more extreme behaviors, we divide our MaNGA galaxies based on the central gradient of their predicted ex-situ stellar mass fraction. We aim to discover if this classification based on the central gradient can actually trace distinct evolution and merging histories, as well as lead to different observational propoerties. For that, we define the radial gradient/slope $s$ as the difference in median ex-situ stellar mass fraction between the inner and intermediate regions of the galaxy: \looseness-2

\begin{equation}
s = \left\langle f_{\mathrm{ex}} \right\rangle_{r < 0.5 R_\mathrm{e}} - \left\langle f_{\mathrm{ex}} \right\rangle_{0.5 R_\mathrm{e} < r < 1 R_\mathrm{e}}
\end{equation}

Based on this slope definition $s$, we categorize the MaNGA galaxies into four distinct classes:

\begin{itemize} 
    \item \textbf{Flat gradient (low ex-situ):} $|s| < 0.05$ and $f_{\mathrm{ex,2D}} < 0.2$ \\
    In-situ dominated galaxies across all radii. This is the most populated class in the MaNGA sample, containing 5909 galaxies ($\sim 58\%$).
    
    \item \textbf{Flat gradient (high ex-situ):} $|s| < 0.05$ and $f_{\mathrm{ex,2D}} > 0.4$ \\Accretion-dominated galaxies across all radii. In our MaNGA sample, this class includes 1237 galaxies ($\sim 12\%$).
    
    \item \textbf{Positive ex-situ gradient:} $s > 0.075$ \\
    Galaxies with an in-situ dominated central region and an ex-situ stellar mass fraction that increases with radius. This class consists of 190 galaxies in our MaNGA sample ($\sim 2\%$).

    \item \textbf{Negative ex-situ gradient:} $s < -0.075$ \\
    Galaxies with a highly accreted central region and an ex-situ stellar mass fraction that decreases outward. This is this class consists of 330 galaxies ($\sim 3\%$).

    \item \textbf{Other/Mixed:} \\
    Galaxies that do not fall into any of the categories above, lying between the defined classes. This group includes 2577 galaxies ($\sim 25\%$).
\end{itemize}

This classification enables us to probe deviations from average radial ex-situ stellar mass fraction profiles and assess links to galaxy evolutionary histories. The right panel of Figure \ref{fig:radial_profiles} presents the median radial profiles for each class, which, by construction, exhibit the expected behaviors with respect to distance from the center.

Figure \ref{fig:slope_groups} displays 30 representative galaxies per class, highlighting clear differences in spatial structure. Flat gradient (low ex-situ) galaxies are dominated by in-situ stars throughout, while flat gradient (high ex-situ) galaxies maintain a uniformly high ex-situ fraction at all radii. Negative ex-situ gradient galaxies exhibit centrally peaked ex-situ fractions that decline outward. Positive ex-situ gradient galaxies show the opposite: low central ex-situ fractions and increasingly ex-situ dominated outskirts, consistent with inside-out growth and late accretion.

\begin{figure*}
\centering
\includegraphics[width=\linewidth]{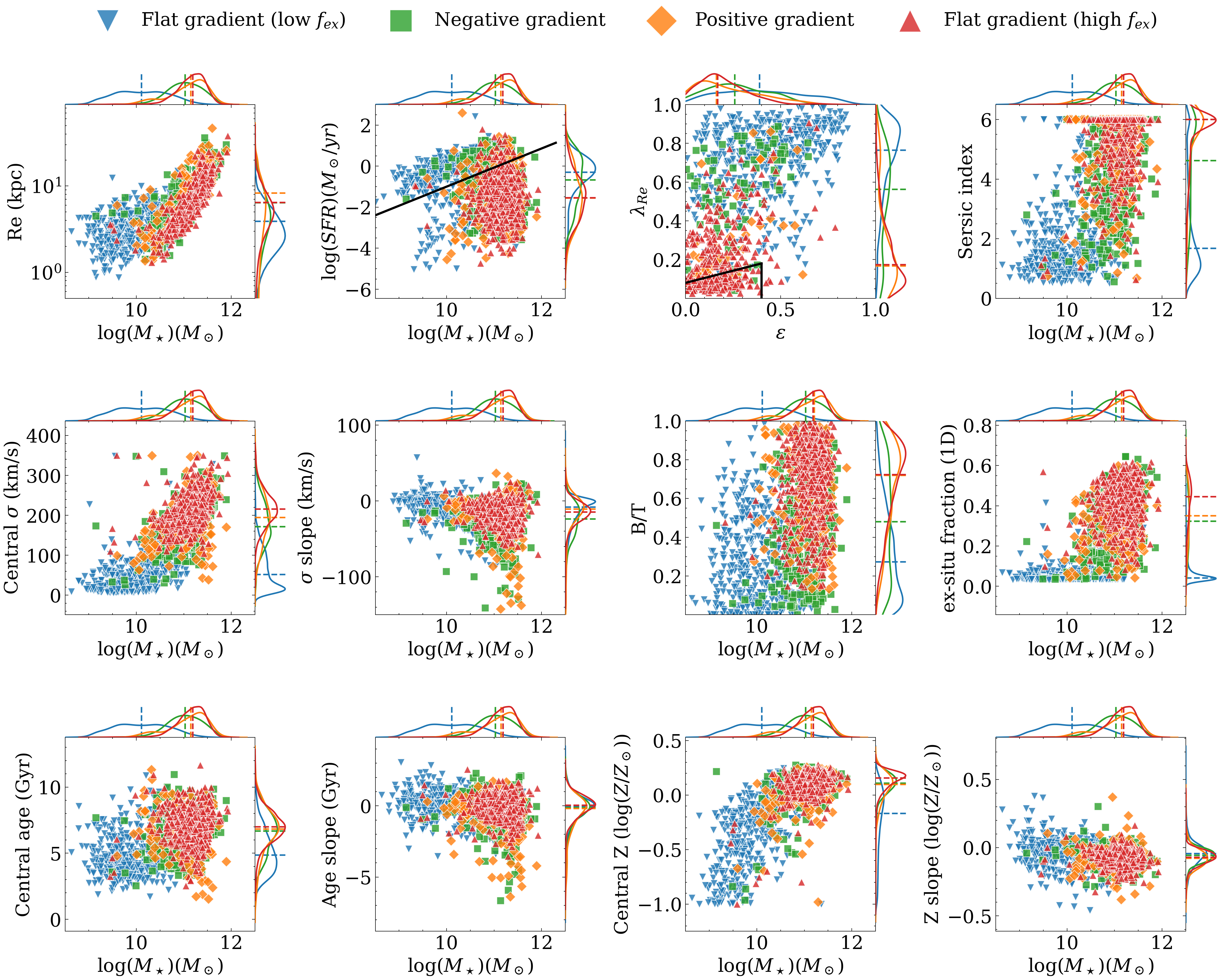}
\caption{Scaling relations of MaNGA galaxies color-coded by ex-situ slope class, showing trends in stellar mass, star formation, rotation, Sersic index, central velocity dispersion, metallicity, age, their gradients, bulge-to-total ratio, and 1D ex-situ fraction (from \citep{2024NatAs...8.1310A}) Classes follow a sequence from low-mass, star-forming, rotationally-supported to high-mass, quenched, dispersion-supported systems.}
\label{fig:scaling_relations}
\end{figure*}

We examine the abundances of each group in each stellar mass bin for our MaNGA sample in the left panel of Fig. \ref{fig:mass_fractions_groups}. Flat gradient (low ex-situ) galaxies prevail in the low stellar mass bin ($\sim$ 84\%) and become less and less common as we move to higher stellar masses.  As the most populous class in our sample, this confirms that the majority of galaxies exhibit flat radial ex-situ profiles. On the contrary, flat gradient (high ex-situ) galaxies are the dominant class in the high stellar mass bin ($\sim$ 39\%). This is expected from the hierarchical scenario of galaxy evolution, as low mass galaxies typically are impacted less from mergers than high mass galaxies. The negative and positive gradient in ex-situ galaxies are less common but are mostly found in intermediate and high stellar mass. 

We further compare these observed abundances to those derived from the MaNGIA mock dataset (middle panel of Fig. \ref{fig:mass_fractions_groups}), focusing especially on the MaNGIA subsample used to train the diffusion model (right panel of Fig. \ref{fig:mass_fractions_groups}).  This comparison can give us insights on the different distributions and diversity between the mock and the observed dataset as well as verify that the model is learning meaningful correlations and not merely mirrors the train set. We find that while the general trends remain consistent - flat-gradient low ex-situ galaxies are more common at low stellar mass and other classes increase in fraction toward higher stellar masses - there are notable differences. Specifically, the flat-gradient (high ex-situ) class is significantly underrepresented in the mock dataset relative to observations, whereas the negative gradient class is more populous in the high stellar mass bin in MaNGIA than in MaNGA. This indicates that while the model successfully captures meaningful correlations learned from the simulations, the observed galaxy population exhibits a different distribution of ex-situ gradient classes, reflecting possible physical diversity between the observations and the simulation.\looseness-2

Apart from the stellar mass dependency, we also examine morphology, star formation activity, kinematics, host dark matter halo mass, and environment across the defined classes (Figure~\ref{fig:slope_group_trends}). Clear and systematic differences emerge. \textbf{Flat gradient (low ex-situ)} galaxies are predominantly disk-dominated, star-forming, fast-rotating systems residing in low-mass halos. \textbf{Negative gradient} galaxies are both late-type and early-type, rotationally supported systems located in intermediate-mass halos ($\log M_\mathrm{halo} < 13$), and include both star-forming and quenched populations. In contrast, 
\textbf{positive gradient} galaxies are primarily quenched ellipticals, both fast and slow rotators, and typically reside in more massive halos ($\log M_\mathrm{halo} > 13$), often in clusters or filaments. 
\textbf{Flat gradient (high ex-situ)} galaxies are the most massive, quenched, early-type galaxies in the sample, commonly found in the most massive halos and in cluster centers. The remaining unclassified galaxies represent a mix of properties. A consistent progression across classes is observed in all examined galaxy properties, underlying that the probed gradient in central ex-situ stellar mass content is tracing diverse groups of galaxies in terms of their characteristics and assembly histories. Notably, these trends persist across different stellar mass ranges, as demonstrated in Appendix~\ref{appendix_mass_trends}.

We further investigate these differences through scaling relations shown in Figure \ref{fig:scaling_relations}, where galaxies are color-coded by class. A clear evolutionary sequence emerges: flat gradient (low ex-situ) galaxies occupy the low-mass, star-forming, rotationally-supported regime, while flat gradient (high ex-situ) galaxies dominate the high-mass, quenched, dispersion-supported regime. The positive and negative gradient classes lie between these extremes. Negative gradient galaxies display a mix of star-forming and quenched systems, whereas positive gradient galaxies are predominantly quenched. Slow rotators are concentrated among the flat gradient (high ex-situ) and positive gradient classes. 

The flat gradient (low ex-situ) class features lower Sersic indices, central velocity dispersions, and bulge-to-total ratios, consistent with disk-dominated, rotationally supported galaxies. These systems tend to have younger stellar ages and lower metallicities, with relatively flat radial gradients. In contrast, the flat gradient (high ex-situ) and positive gradient classes exhibit higher Sersic indices, larger central velocity dispersions, and more prominent bulges, reflecting their spheroidal, dispersion-supported nature. These classes also host older, more metal-rich stellar populations. The negative gradient class typically occupies an intermediate position, with properties bridging those of the low and high ex-situ slope classes, showing mixed characteristics. \looseness-2

These systematic variations across classes highlight distinct assembly histories and evolutionary paths linked to the spatial distribution of accreted stellar material. Overall, the classes defined by the central ex-situ fraction slopes correspond closely to distinct galaxy characteristics and evolutionary stages. Many trends align with the hierarchical evolution model, while others are less intuitive and will be further explored in the next section.

\subsection{Recovery of classes in MaNGIA}

\begin{figure*}
    \centering
    \includegraphics[width=\linewidth]{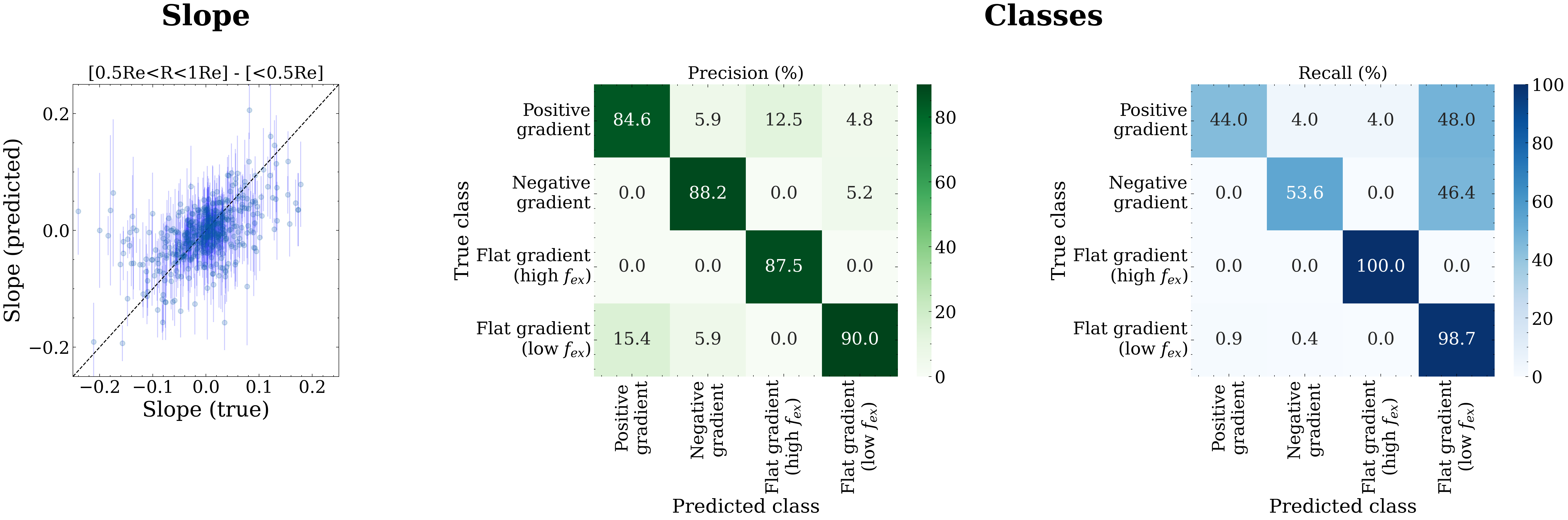}
    \caption{Recovery of the slope and the classes classification in the MaNGIA test set. Left: comparison between the true and predicted slope in the MaNGIA test set. The one-to-one line is shown for reference. While the model recovers the overall trends, extreme slope values are more difficult to capture. The error bars correspond to the standard deviation of the recovered slopes from the 100 samples. Right: confusion matrix for the classification on the same test set. Precision is high in all classes ($>84\%$), meaning false positives are rare. Recall is lower for the underrepresented positive and negative slope classes, which are often classified as the most common flat-slope (low ex-situ) class.}
    \label{fig:slope_recovery}
\end{figure*}

Since much of the analysis relies on the slope of the ex-situ stellar mass fraction in the inner regions, as well as the resulting classification, it is important to test how well both quantities are recovered in the MaNGIA test set. Such an evaluation is only possible with the mock dataset, where the ground truth can be directly calculated. In Fig.~\ref{fig:slope_recovery}, we present both the one-to-one recovery of slope values and the confusion matrix for the MaNGIA test set.

We find that recovering the slope is more difficult than the global and regional metrics shown in Fig.~\ref{fig:model_metrics}, particularly in the most extreme cases. This is expected, since slope recovery was not a primary training target of the diffusion model but a secondary diagnostic introduced for the science analysis. Nevertheless, the predicted slopes broadly follow the one-to-one relation, with larger scatter and uncertainties at the extremes.

We also test how well the four slope classes can be recovered in the MaNGIA test set. To this end, we calculate the confusion matrix, shown in the right panel of Fig.~\ref{fig:slope_recovery}, and evaluate both precision and recall. Precision measures how often a predicted class is correct. We find that it is high across all classes ($>84\%$), meaning that when the model assigns a class, it is generally reliable and false positives are rare. Recall, on the other hand, measures how many of the true cases are successfully recovered. Here the model performs less well for the positive and negative slope classes, which are underrepresented in the training data. These cases are frequently classified as the most common flat-slope (low ex-situ) class, a typical outcome in highly imbalanced classifications.

Overall, while the classification is not perfect, precision remains very high across all classes. This is the main objective of the analysis, as we emphasize that the diffusion model was not trained specifically for this task; classification was introduced later as a secondary step motivated by the analysis of the recovered results. A more tailored approach would likely improve performance further, but the diffusion model provides higher-order information beyond what a simple classifier can capture. For the statistical goals of this work, the current recovery is sufficient.

\section{Discussion}\label{sec:discussion}

\subsection{Connection with Merger Histories from TNG50}

\begin{figure*}
    \centering
    \includegraphics[width=\linewidth]{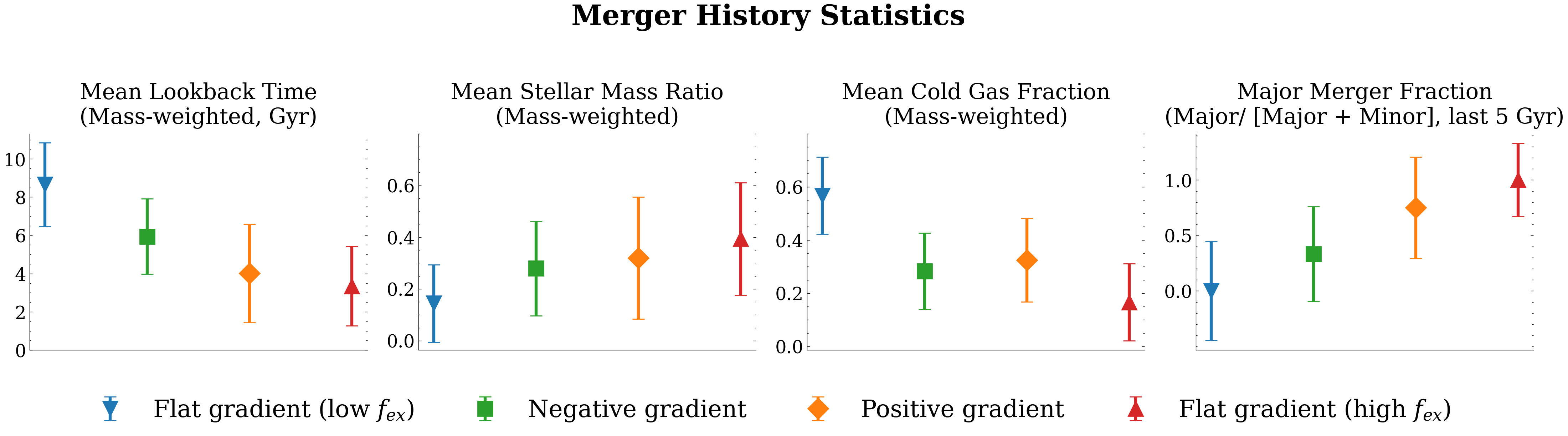}
    
    \caption{Distributions of key merger history metrics for MaNGIA galaxies, separated by ex-situ slope class. We plot the mean lookback time of mergers, weighted by the stellar mass of the secondary progenitors, reflecting when most mergers occurred; the mean cold gas fraction of merging companions, weighted by their stellar mass, indicating the typical gas richness of the merger history; the mean stellar mass ratio of mergers, also weighted by secondary progenitor mass, characterizing the relative strength of merger events; the fraction of major mergers, calculated as the number of major mergers divided by the total number of major and minor mergers over the last 5 Gyr, highlighting the dominance of major versus minor mergers in recent assembly history. We plot the median value and the error bars correspond to the 68\% of the data.}

    \label{fig:merging_history_properties}
\end{figure*}

We aim to investigate whether the galaxy classes identified based on their inferred ex-situ stellar mass fraction radial profiles indeed correspond to different evolution histories. To this end, we turn to the TNG50 cosmological simulation, from which our MaNGIA training set is derived. In TNG50, we have access to the full merging histories of simulated MaNGIA galaxies, allowing us to directly connect the final observed properties to their past evolution. We apply the same classification scheme introduced previously to the full MaNGIA dataset. As in the observational sample, the most populous class is the flat gradient (low ex-situ) galaxies, comprising approximately half of the total sample. The abundances of each group separated by stellar mass can be found in the middle panel of Fig. \ref{fig:mass_fractions_groups}.

In Figure \ref{fig:merging_history_properties}, we plot the distribution of several key metrics that summarize the merging histories of MaNGIA galaxies, grouped by their ex-situ slope classification. These metrics are derived from TNG50 \citep{2017MNRAS.467.3083R, 2023MNRAS.519.2199E}. Specifically, we show: (1) the mean lookback time of mergers, weighted by the stellar mass of the secondary progenitors, which traces when most merger activity occurred; (2) the mean cold gas fraction of merging companions, similarly mass-weighted, which serves as a proxy for the typical gas richness or “wetness” of a galaxy’s accretion history; (3) the mean stellar mass ratio of mergers, indicating the strength of these interactions; and (4) the recent fraction of major mergers, calculated as the number of major mergers divided by the total number of major and minor mergers in the last 5 Gyr, offering a measure of the most frequent recent merger type.

We find clear distinctions between the classes. Flat gradient (low ex-situ) galaxies exhibit the oldest merger lookback times, the highest mean cold gas fractions, and the lowest mean stellar mass ratios and recent major merger fractions -supporting a quiescent, gas-rich accretion history dominated by minor mergers. In contrast, flat gradient (high ex-situ) galaxies show the youngest merger lookback times, the highest stellar mass ratios, and the lowest cold gas fractions, consistent with a recent history dominated by dry, major mergers. Negative gradient galaxies present intermediate properties, with relatively early, moderately gas-rich merger activity. This supports the scenario of an early on major merger followed with a more quiescent history. Positive gradient galaxies tend to experience more recent and gas-poor mergers, with elevated stellar mass ratios and a higher fraction of recent major mergers. The key difference between the two most accretion-dominated classes - flat gradient (high ex-situ) and positive gradient galaxies - is that the former appears to have undergone somewhat more recent, drier mergers, with a higher major-to-minor merger ratio within the last 5 Gyr.

These results strongly support the interpretation that the differences observed in the present-day ex-situ radial profiles are imprints of different merger-driven evolution histories, encoded in the stellar populations of galaxies.  Notably, these trends hold across different stellar mass ranges, as shown in Appendix~\ref{appendix_merging_stats_mass}. A more detailed view of the stacked merger histories for each group is presented in Appendix~\ref{merger_histories_appendix}, highlighting how the ex-situ stellar mass fraction evolves differently over time across the four classes.

\subsection{Evolution Scenarios}

\begin{figure*}
    \centering
    \includegraphics[width=\linewidth]{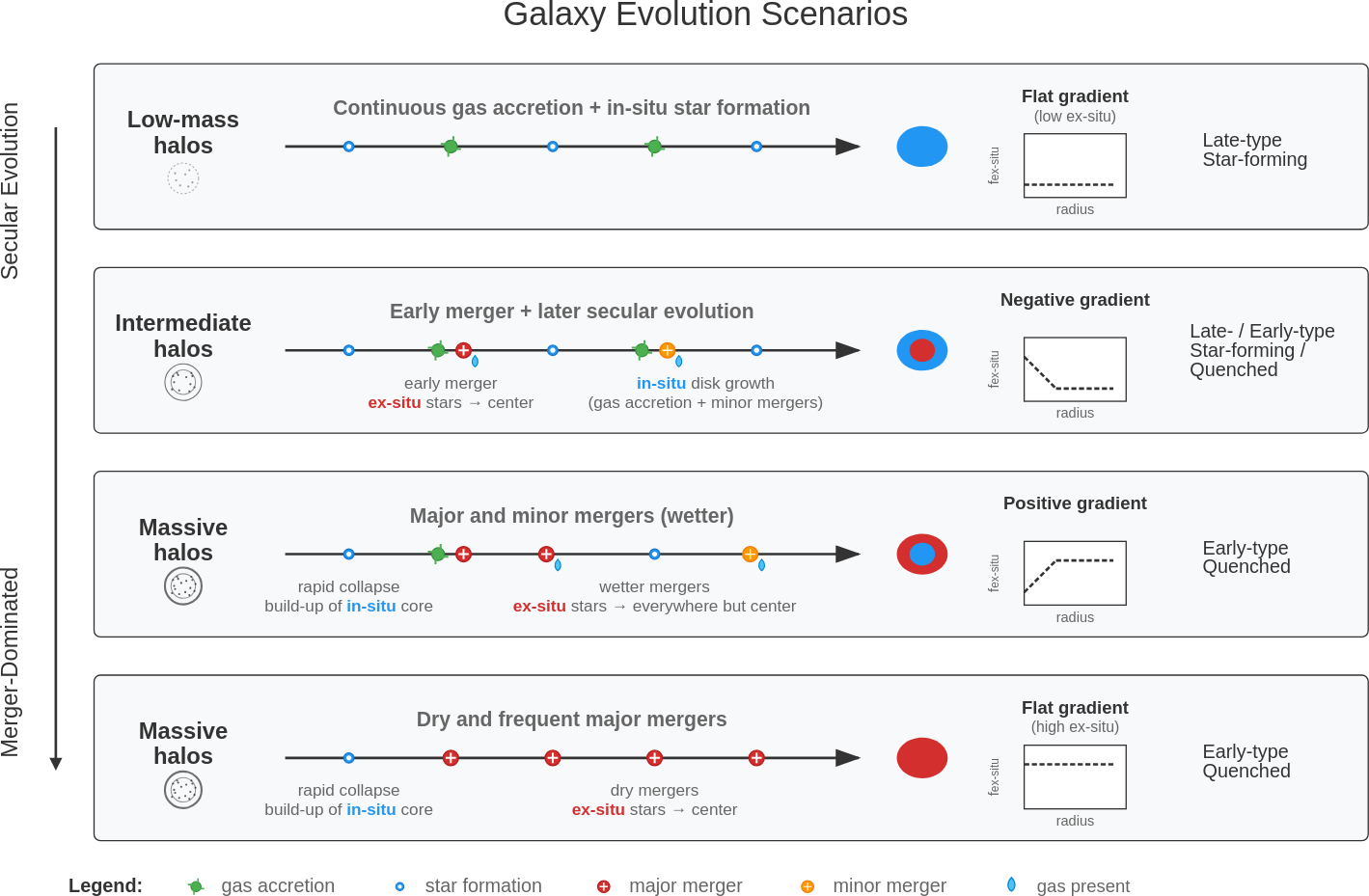}
    \caption{Illustration of  the proposed evolutionary scenarios corresponding to the four classes defined by central ex-situ stellar mass fraction gradients. The sequence spans from secularly evolving to merger-dominated systems.}
    \label{fig:scenarios}
\end{figure*}

In the previous section, we classified MaNGA galaxies into four categories based on the gradients in their radial ex-situ stellar mass fraction profiles within the central regions. We explored how these classes differ in terms of physical and morphological properties, scaling relations, and merging histories using the TNG50 cosmological simulation, from which our MaNGIA training set originates. Based on these comparisons, we propose four distinct evolutionary scenarios that give rise to the observed diversity in the central stellar mass assembly of galaxies. A schematic representation of these scenarios is presented in Figure~\ref{fig:scenarios}.

Across several scaling relations there is some level of overlap among the different classes, starting with \textbf{flat gradient (low ex-situ)} typically on one end and leading to \textbf{flat gradient (high ex-situ)} on the other, with the remaining two classes in between. This trend likely reflects a physical shift from galaxies with largely secular, in-situ star formation histories to systems whose stellar mass assembly is dominated by mergers. 

We begin by analyzing our findings on the \textbf{flat gradient (low ex-situ)} class. We find that these systems tend to be young, metal-poor, star-forming (76\%), with late-type morphologies (87\%) and fast-rotating kinematics (98\%). Their flat and uniformly low ex-situ radial profiles indicate an in-situ dominated assembly at all radii. In TNG50, the simulated analogs show quiet merger histories and continuous in-situ star formation. Observationally, this is consistent with well-established models of secular disk evolution in low-mass galaxies \citep{2004ARA&A..42..603K, 2014RvMP...86....1S}. The environments in which these galaxies reside may also have influenced their stellar mass assembly. These systems are typically found in low-mass halos (61\%), below the threshold for virial shock heating \citep{2006MNRAS.368....2D}, $M_{\mathrm{shock}} < 10^{12} M_\odot$. In this regime, cold gas can continuously accrete onto the galaxy, supporting ongoing star formation.

In contrast, the \textbf{negative gradient} class suggests a more complex assembly history. These galaxies show high ex-situ stellar mass fractions in their centers that decline with radius. They have both late-type (42\%) and early-type morphologies (46\%) and fall within an intermediate to high stellar mass range ($\sim 10^{11} M_\odot$). Notably, this class includes both star-forming (35\%) and quenched (65\%) populations. On average, their central regions host older stellar populations and higher velocity dispersions, supporting a two-phase formation model. In TNG50, such profiles emerge from early gas-rich major mergers - typically around 7–10 Gyr ago - that lead to a compact, accreted bulge, followed by a prolonged phase of disk regrowth through minor mergers and in-situ star formation. However, we emphasize that such cases represent only a small fraction of galaxies in TNG. Previous studies have shown that the majority of bulges, especially in galaxies with $M_\star < 10^{11} M_\odot$, formed predominantly through in-situ processes, with mergers playing only a minor role in their buildup \citep{2025A&A...699A.320Z}. Results from N-body simulations support that high ex-situ mass fractions in the center can emerge from mergers early in the history of the galaxy \citep{2017MNRAS.464.2882A}. The idea that bulges can be largely assembled through early mergers, while disks form later through smoother processes, is also supported by observations \citep{2021MNRAS.504.3058M, 2025NatAs...9..141B}. The late-type morphology that appears in this class is consistent with inside-out disk growth via minor accretion and extended star formation \citep{2014ApJ...788...28V}. Galaxies in this class typically reside in intermediate-mass halos ($M_{halo} < 10^{13}$, 58\%), close or slightly above the $M_{shock}$ threshold. This suggests that cold gas can still penetrate the halo and trigger star-formation, further supporting the build-up of an in-situ disk from gas accretion along with minor mergers, and possibly explaining the presence of both star-forming and quenched galaxies within this class.

The \textbf{positive gradient} and \textbf{flat gradient (high ex-situ)} classes are the two classes more closely related in terms of stellar mass and environment. They both have high ex-situ stellar mass fractions in their outer regions but differ in their central parts: positive gradient galaxies have an in-situ dominated core, whereas flat gradient (high ex-situ) galaxies are ex-situ dominated at all radii. These are the most merger-dominated systems and are typically hosted by more massive halos, where gravitational collapse is more rapid, enabling an early and compact in-situ core to form. Galaxies in both classes may have followed a 'blue nugget' scenario \citep{2014ApJ...791...52B, 2015MNRAS.450.2327Z}, where violent disk instabilities, triggered by mergers or gas inflows (common in high-density environments), cause intense central starbursts and bulge formation \citep{2023MNRAS.522.4515L}. However, their subsequent growth diverges depending on the accretion history and cold gas availability, shaping their final radial ex-situ profiles. 

Galaxies in the \textbf{positive gradient} mainly occupy the high end of the stellar mass range ($M_\star> 10^{11} M_\odot$) and are primarily early-type (81\%) and quenched systems (84\%). Their positive ex-situ gradients, coupled with bulge-disk structural distinctions, align with the predictions of two-phase formation models, where early central buildup is followed by late accretion \citep{2016MNRAS.458.2371R, 2020MNRAS.497...81D}. The high ex-situ stellar mass fractions beyond the in-situ-dominated center suggest that these galaxies have undergone mergers with substantial dynamical impact — though not with companions massive enough to reach and dominate the core \citep{2017MNRAS.464.2882A}. In TNG50, galaxies in this class indeed show evidence of a more active merger history, with relatively gas-rich (\textit{wet}) mergers, which can permit rejuvenation of central star formation even after the onset of shock heating (as seen in their MaNGIA merger histories in Fig. \ref{fig:merging_history_properties}). This process can preserve or regenerate an in-situ core while ex-situ stars accumulate predominantly in the outskirts.  While these galaxies typically reside in massive halos above the $M_{\mathrm{shock}} > 10^{12} M_\odot$ threshold (88\%), cold flows at high redshift may have further sustained star formation in the center by delivering gas along cosmic web filaments \citep{2006MNRAS.368....2D, 2013MNRAS.435..999D}. Meanwhile, repeated minor mergers built up the ex-situ-rich outskirts, producing the observed positive radial gradient in ex-situ fraction.

Finally, galaxies in the \textbf{flat gradient (high ex-situ)} class are among the most massive in the sample ($M_\star> 10^{11} M_\odot$) and are predominantly early-type (89\%) and quenched (89\%). Their high ex-situ content and quenched nature are consistent with previous studies linking rapid quenching to mergers that trigger intense starbursts and AGN activity, ultimately depleting the cold gas reservoir \citep{2018MNRAS.473.1168R, 2019ApJ...874...17B}. Additionally, these systems typically reside in massive halos above the $M_{\mathrm{shock}} > 10^{12} M_\odot$ threshold (93\%), where sustained cold gas inflows are suppressed, further contributing to quenching. In TNG50, these galaxies exhibit a violent merger history, dominated by frequent, massive dry mergers. Such events quench central star formation and build up a stellar structure that is almost entirely accreted. Unlike the positive gradient systems, their merging companions were massive enough to penetrate into the inner regions, depositing ex-situ stars even in the core \citep{2017MNRAS.464.2882A}. As a result, they display uniformly high ex-situ mass fractions from the center to the outskirts. These systems likely evolved from compact, early-formed progenitors that grew substantially in size through dry accretion — a process consistent with the \textit{red nugget} scenario \citep{2006MNRAS.373L..36T, 2008ApJ...677L...5V}, where dense cores formed at high redshift expand over time via minor and major dry mergers \citep{2009ApJS..182..216K, 2015ApJ...813...23V}. The lack of cold gas inflows prevents any rejuvenation of central star formation, resulting in the most merger-built stellar structures in our sample.

A key driver behind these divergent evolutionary paths may be the galaxy environment. In particular, halo mass has been shown to influence stellar population properties \citep{2024NatAs...8..648S, 2024ApJ...974...29O}, and likely plays a central role in shaping stellar mass assembly. Environment affects both the frequency and type of mergers, as well as the availability of cold gas. Previous studies have found that higher merger rates—especially those involving dry mergers—are more frequent in overdense environments and galaxy groups \citep{2010ApJ...718.1158L, 2012ApJ...754...26J}, while gas accretion is regulated by halo mass and the onset of virial shock heating \citep{2006MNRAS.368....2D}. Together, these factors help explain the observed sequence of ex-situ gradients across our classes — from in-situ dominated disks in low-mass halos to ex-situ dominated ellipticals in massive halos —and are highly consistent with the results presented in this work.

\subsection{Similar classifications in simulations}
The four classes identified in this work exhibit strong similarities with equivalent classes proposed for early-type galaxies (ETGs) in previous works on the IllustrisTNG (TNG100) \citep{2021A&A...647A..95P} and Magneticum \citep{2022ApJ...935...37R} simulations. Both of these studies developed classification schemes based on the radial distribution of ex-situ stellar mass, although they primarily targeted more massive systems and extended their analysis to significantly larger radii (up to $\sim 10R_e$). Despite these differences in sample and spatial coverage, we observe a consistent correspondence in the types of profiles identified. In particular, both studies report an in-situ dominated class — less populous in their samples due to the focus on ETGs — and a fully ex-situ dominated class, analogs to our \textbf{flat gradient (low ex-situ)} and \textbf{flat gradient (high ex-situ)} groups, respectively. Their most populous profile closely resembles our \textbf{positive gradient} class, although defined for larger galactic extents. An equivalent to our \textbf{negative gradient} class is also present in both works, including also galaxies that show an ex-situ dominated core transitioning to an in-situ dominated intermediate region, and then back to ex-situ dominance in the outskirts. While we focus exclusively on the central regions, extending our analysis to larger radii would likely reveal similar three-part structures.

These consistencies suggest that the physical mechanisms driving radial variations in stellar mass origin — such as merger timing, mass ratios, and gas content — are robust across mass scales and simulations. Moreover, the merger statistics presented in \citet{2021A&A...647A..95P} closely mirror those we observe in TNG50, reinforcing the validity and broader relevance of the evolutionary pathways proposed in this study. Overall, these four classes appear to trace a continuum in galaxy assembly histories, from secularly evolving disk galaxies to massive, merger-built ellipticals. Furthermore, the stacked star formation histories for each group, derived from the Firefly value-added catalog \citep{2022MNRAS.513.5988N, 2017MNRAS.466.4731G} and presented in Appendix \ref{appendix_sfhs}, further support the proposed evolutionary scenarios.

\subsection{Limitations and Caveats}

While our approach demonstrates strong performance in reconstructing ex-situ stellar mass fraction maps, several limitations and caveats remain. First, the size of the training sample (MaNGIA) is relatively small, primarily due to the computational cost of forward-modelling mock observables from cosmological simulations. This constraint limits the diversity of physical conditions represented in the training set and may affect the model’s ability to generalize to galaxies that might not be present in the training distribution. While we found that such cases are not so common, two examples can be found in Appendix ~\ref{appendix_failed}.

Additionally, the current diffusion model is conditioned on only three observable maps — stellar mass density, velocity, and velocity dispersion — each normalized to highlight spatial gradients. While this input is physically meaningful, it may not fully suffice to predict the diverse ex-situ stellar mass fraction distribution. Incorporating additional features or data of different modalities could further improve the model’s ability to infer the spatial structure of the ex-situ component.

Most importantly, the current methodology is trained on a particular mock dataset (MaNGIA), derived from the cosmological simulation TNG50. Although we have demonstrated in our previous works that the used conditions, namely the stellar mass density and kinematic maps, are robust predictors of the ex-situ stellar mass fraction across different simulations \citep{2023MNRAS.523.5408A, 2024NatAs...8.1310A}, extending this framework to include a broader range of cosmological models and subgrid physics would offer valuable insights into the model's generalizability. At present, such an expansion is limited by the availability of mock data for IFU surveys. However,  a promising direction could involve adopting a foundation model approach \citep{2024MNRAS.531.4990P, 2025arXiv250315312E}, in which a model is pretrained on a large, heterogeneous dataset including multiple simulations, forward-modeled outputs, and observational data. This strategy could reduce dependence on expensive forward-modelling, enhance predictive robustness, and enable broader applicability across obsevrational datasets and redshifts.

\section{Conclusion}\label{sec:conclusion}

We predict two-dimensional, spatially-resolved ex-situ stellar mass fraction maps for 10,000 galaxies in MaNGA using a diffusion model trained on mock MaNGA analogs (MaNGIA), which are in turn obtained from the TNG50 cosmological simulation. Our main findings are summarized below:

\begin{itemize}
    \item The diffusion model accurately recovers the global ex-situ stellar mass fraction when aggregating the predicted 2D maps, as well as localized features within different apertures. These predictions are based solely on spatial gradients in stellar mass, velocity, and velocity dispersion maps, and are validated on an unseen test set from MaNGIA.

    \item When applied to MaNGA galaxies, the model generally predicts flat ex-situ stellar mass fraction radial profiles. In low-mass galaxies, these profiles are centered around low ex-situ fractions and show little to no radial dependence between the central regions and larger radii (e.g., 1.5$R_e$ to 2.5$R_e$). In more massive systems, the global ex-situ fraction increases, and a mild radial gradient appears, with central regions exhibiting median ex-situ fractions approximately $\sim$0.1 lower than at 2.5$R_e$.

    \item We classify galaxies based on the slope of their predicted ex-situ stellar mass fraction profile between the central ($r < 0.5R_e$) and intermediate ($0.5R_e < r < 1R_e$) regions. This classification reveals four distinct assembly modes that correlate with galaxy morphology, mass, star formation activity, and merger history. These categories are the following in terms of their ex-situ stellar mass fraction in the central regions: flat gradient (low ex-situ), negative gradient, positive gradient, and flat gradient (high ex-situ). The 4 distinct classes appear to link spatially-resolved ex-situ content with different evolutionary pathways.
    
    \item We verify these different merging histories in simulation space by applying the same classification to MaNGIA galaxies, finding consistent trends in merger frequency, mass ratios, and timing across the four ex-situ profile classes.

    \item Across the classified populations, we observe a clear separation depending on galaxy assembly histories — from secularly evolving, low-mass star-forming disks to massive, merger-built ellipticals — with the slope of the ex-situ profile serving as a powerful diagnostic of past accretion processes.

\end{itemize}

These results demonstrate the ability of generative models to extract detailed formation histories from present-day galaxy kinematics, offering a new path to spatially resolving assembly histories in large galaxy surveys. In the near future, we plan to extend this methodology to extract even more direct insights into galaxies’ merger histories, with the goal of applying it across a broad range of IFU surveys spanning diverse redshifts and environments.

\begin{acknowledgements}
We thank the referee for their constructive comments which improved the clarity of the paper. EA thanks Glenn van de Ven and L. Scholz-D\'iaz for their insightful discussions.  MHC and EA acknowledge financial support from the State Research Agency (AEI\-MCINN) of the Spanish Ministry of Science and Innovation under the grants ``Galaxy Evolution with Artificial Intelligence" with reference PGC2018-100852-A-I00 and "BASALT" with reference PID2021-126838NB-I00. JFB and EA acknowledge support the Spanish Ministry of Science, Innovation and Universities (Grant Nos. PID2019-107427GB-C32 and PID2022-140869NB-I00) and through the project TRACES from the Instituto de Astrofísica de Canarias,  which is partially supported through the state budget and the regional budget of the Consejer\'ia de Econom\'ia, Industria, Comercio y Conocimiento of the Canary Islands Autonomous Community. This work was supported by Mitacs through the Mitacs Globalink Research Award. This research was enabled in part by support provided by the Digital Research Alliance of Canada. This research also made use of computing time available on the high-performance computing systems at the Instituto de Astrofisica de Canarias. The authors wish to acknowledge the contribution of Teide High-Performance Computing facilities to the results of this research. TeideHPC facilities are provided by the Instituto Tecnológico y de Energías Renovables (ITER, SA). The author thankfully acknowledges the technical expertise and assistance provided by the Spanish Supercomputing Network (Red Espanola de Supercomputacion), as well as the computer resources used: the Deimos-Diva Supercomputer, located at the Instituto de Astrofisica de Canarias.     
\end{acknowledgements}

%
%

\bibliographystyle{aa}
\bibliography{bibliography}

\begin{appendix}

\section{Global ex-situ and in-situ stellar mass trends}
\label{appendix_recreate}

\begin{figure}
    \centering
    \includegraphics[width=\linewidth]{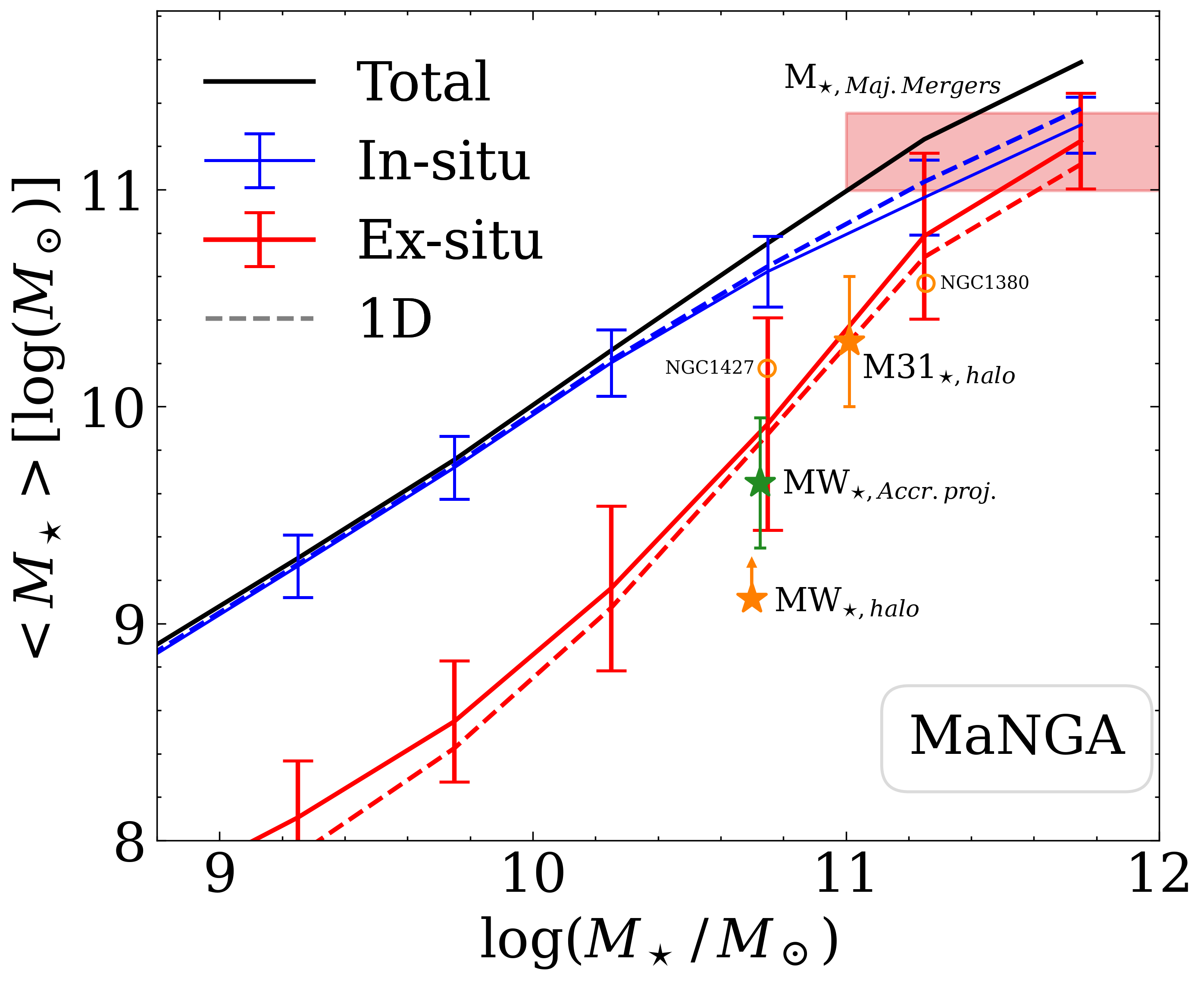}
   
    \caption{Recreation of Fig. 1 from \cite{2024NatAs...8.1310A}, showing the average contributions of ex-situ and in-situ stellar mass for 10,000 MaNGA galaxies in different stellar mass bins, as predicted by the diffusion model. Solid lines indicate the new measurements derived from the aggregated 2D predicted maps. For reference, the previous 1D predictions from \cite{2024NatAs...8.1310A} are shown with dashed lines. The main trends remain consistent between the two works and are in agreement with independent studies of individual galaxies.}
    \label{fig:recreation}
\end{figure}

We recreate Figure 1 from our previous work \citep{2024NatAs...8.1310A}, showing the average contribution of the in situ vs. ex-situ stellar mass in each stellar mass bin for 10,000 MaNGA galaxies using the spatially-resolved maps of ex-situ stellar mass fraction as predicted from the diffusion model in this work. The result is shown in Fig. \ref{fig:recreation} and serves as a validation check of the diffusion model’s predictive capability when extended from the global ex-situ stellar mass fractions presented in our previous study \citep{2024NatAs...8.1310A} to full 2D spatial maps. 

\section{Aperture effects: global ex-situ fraction vs 2D aggregation}
\label{appendix_aperture}

\begin{figure}
    \centering
    \includegraphics[width=0.9\linewidth]{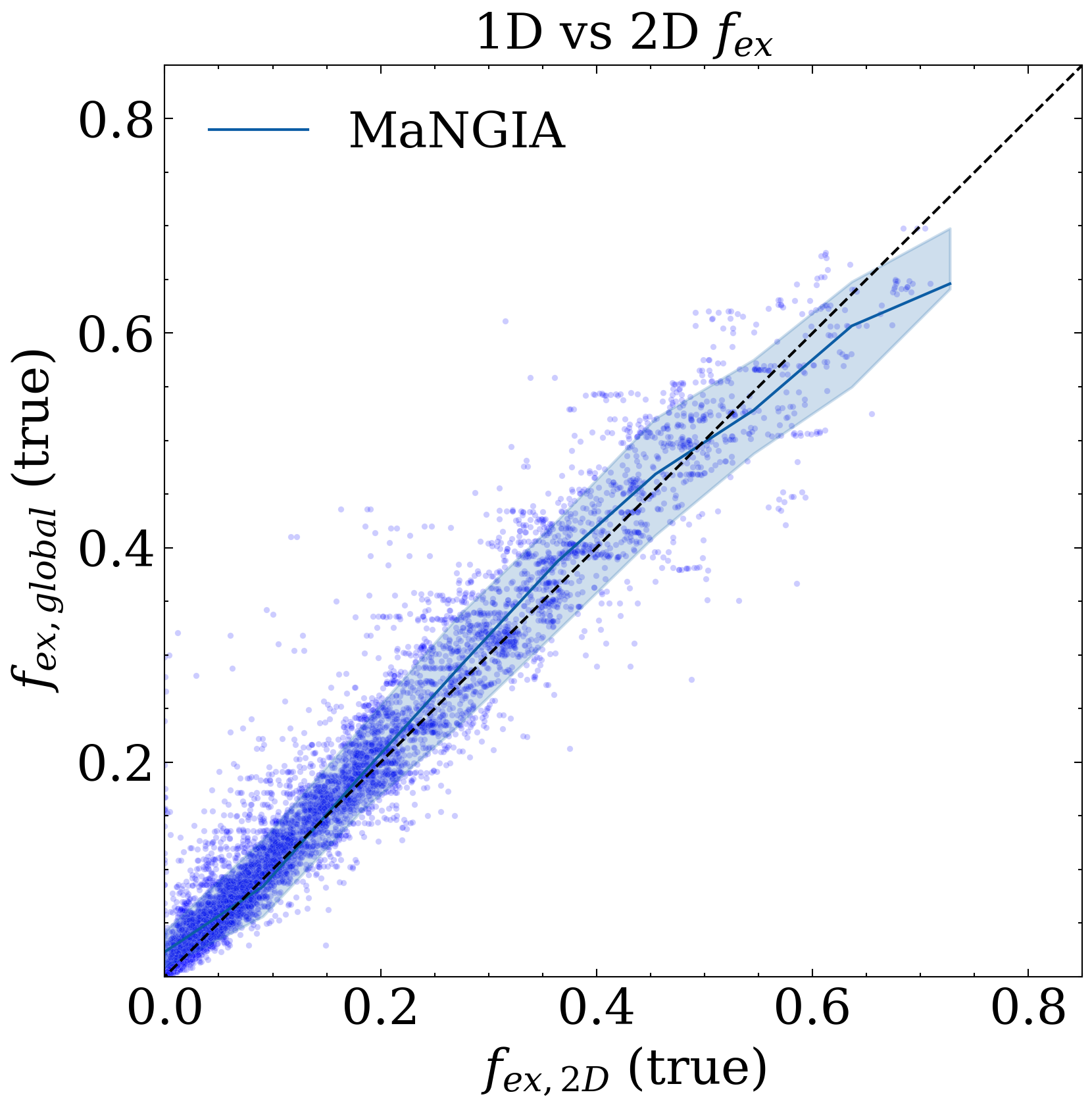}
   
    \caption{Comparison between the true global ex-situ stellar mass fraction and the true ex-situ fraction restricted to the MaNGIA coverage (1.5Re/2.5Re). Each point corresponds to one MaNGIA galaxy. The dashed line indicates the one-to-one relation. The close agreement demonstrates that aperture effects have only a minor impact on the recovered global ex-situ fractions.}
    \label{fig:aperture}
\end{figure}

To check how aperture coverage affects the recovered ex-situ mass, we compare the ground-truth ex-situ fraction measured within the MaNGIA coverage (1.5Re or 2.5Re) to the true global ex-situ fraction from all stellar particles in each galaxy. As shown in Fig. \ref{fig:aperture}, the two quantities are very close. This is because the MaNGIA aperture already encloses the majority of the stellar mass, and regions outside the aperture contribute relatively little additional ex-situ mass. Aperture effects therefore do not strongly change the integrated values used in our analysis.

\section{Failed reconstruction cases}
\label{appendix_failed}

\begin{figure}
    \centering
   \includegraphics[width=\linewidth]{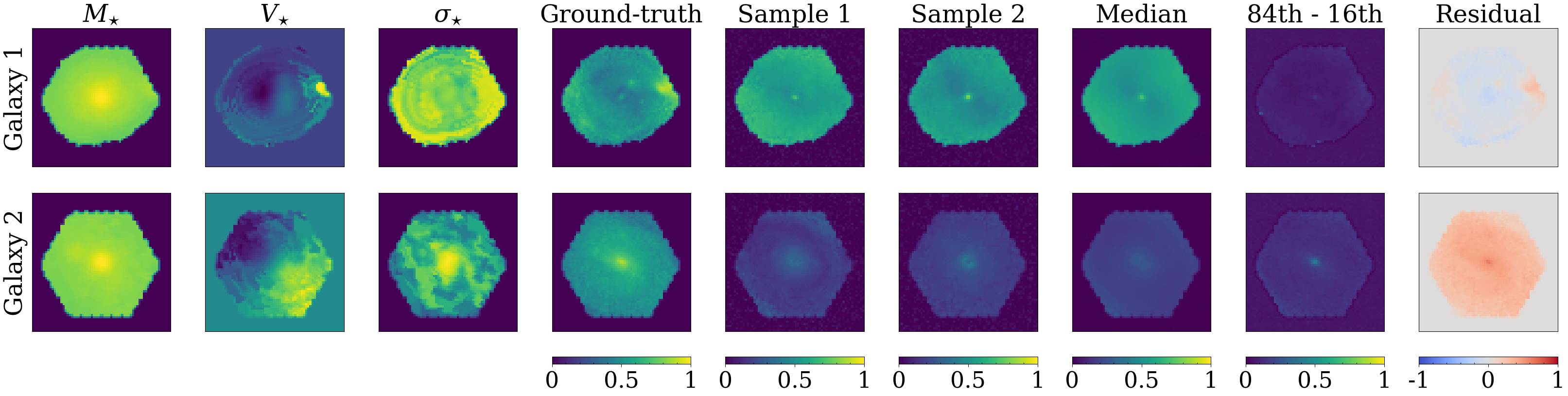}\\
    \caption{Failed reconstruction of the ground-truth from the diffusion model for 2 mock galaxies from the MaNGIA test set.}
    \label{fig:reconstruction_examples_failed}
\end{figure}

We present two failed reconstruction cases from our diffusion model in Fig.~\ref{fig:reconstruction_examples_failed}. In the first row, we show a case where the model successfully captures the large-scale structure but misses finer details, particularly in regions with higher ex-situ fractions. In the second row, we present a completely failed reconstruction, where the input maps do not provide sufficient information for the model to reconstruct the ex-situ stellar mass fraction distribution, resulting in larger residuals. While such cases are rare, they highlight the inherent limitations of working with a relatively small training dataset.
 
\section{MaNGA group trends by stellar mass}
\label{appendix_mass_trends}
\begin{figure*}
    \centering
    \includegraphics[width=\linewidth]{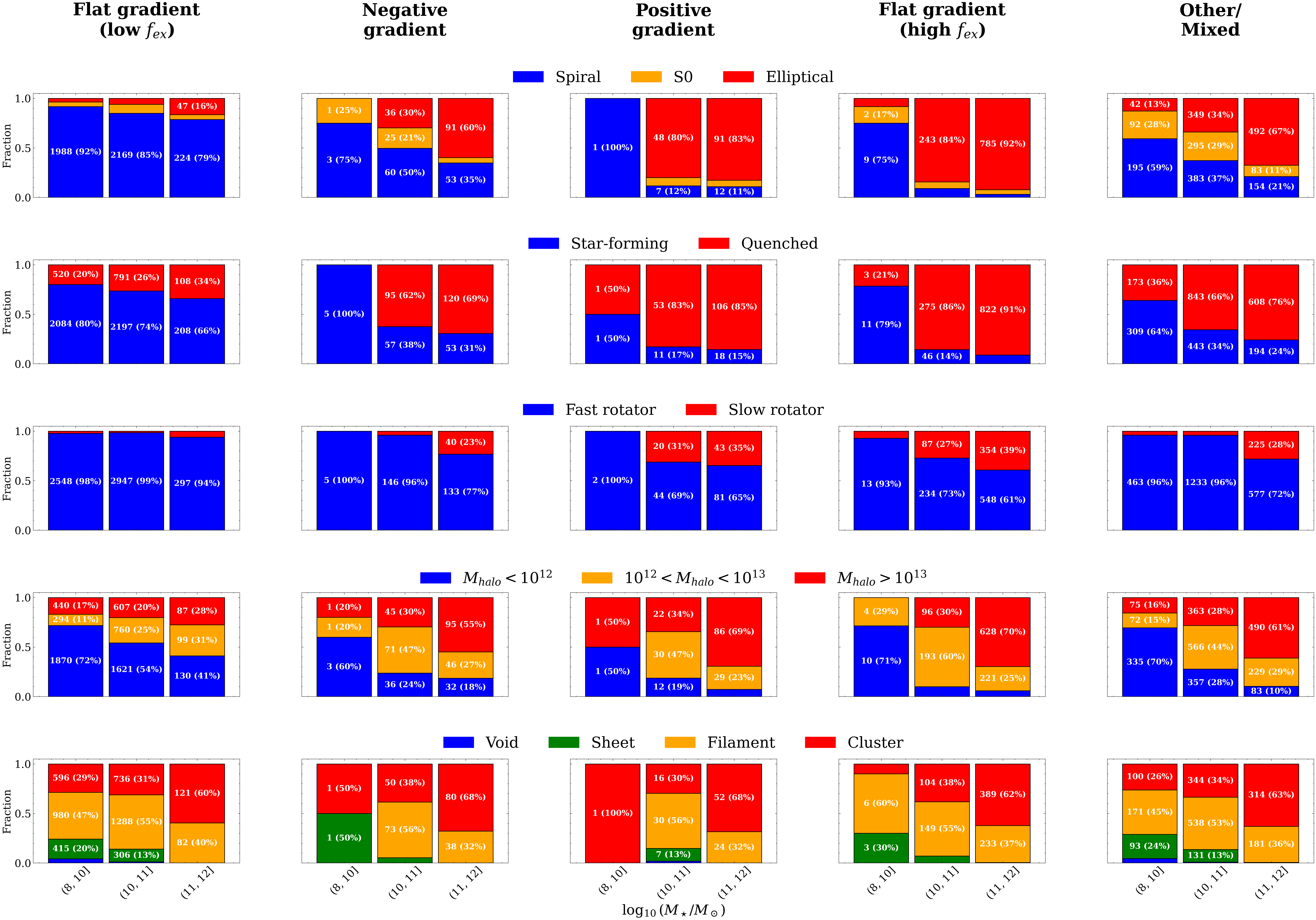}
    \caption{Galaxy properties across ex-situ slope classes, showing clear differences in morphology, star formation, rotation, and environment, also separated by stellar mass.}
    \label{fig:slope_group_trends_mass_bins}
\end{figure*}

We recreate Fig. \ref{fig:slope_group_trends} from the main text, separating in stellar mass bins per category. We find that all trends still hold independent of the probed stellar mass, as shown in Fig. \ref{fig:slope_group_trends_mass_bins}.

\section{MaNGIA merging stats by stellar mass} 
\label{appendix_merging_stats_mass}
\begin{figure*}
    \centering
    \includegraphics[width=\linewidth]{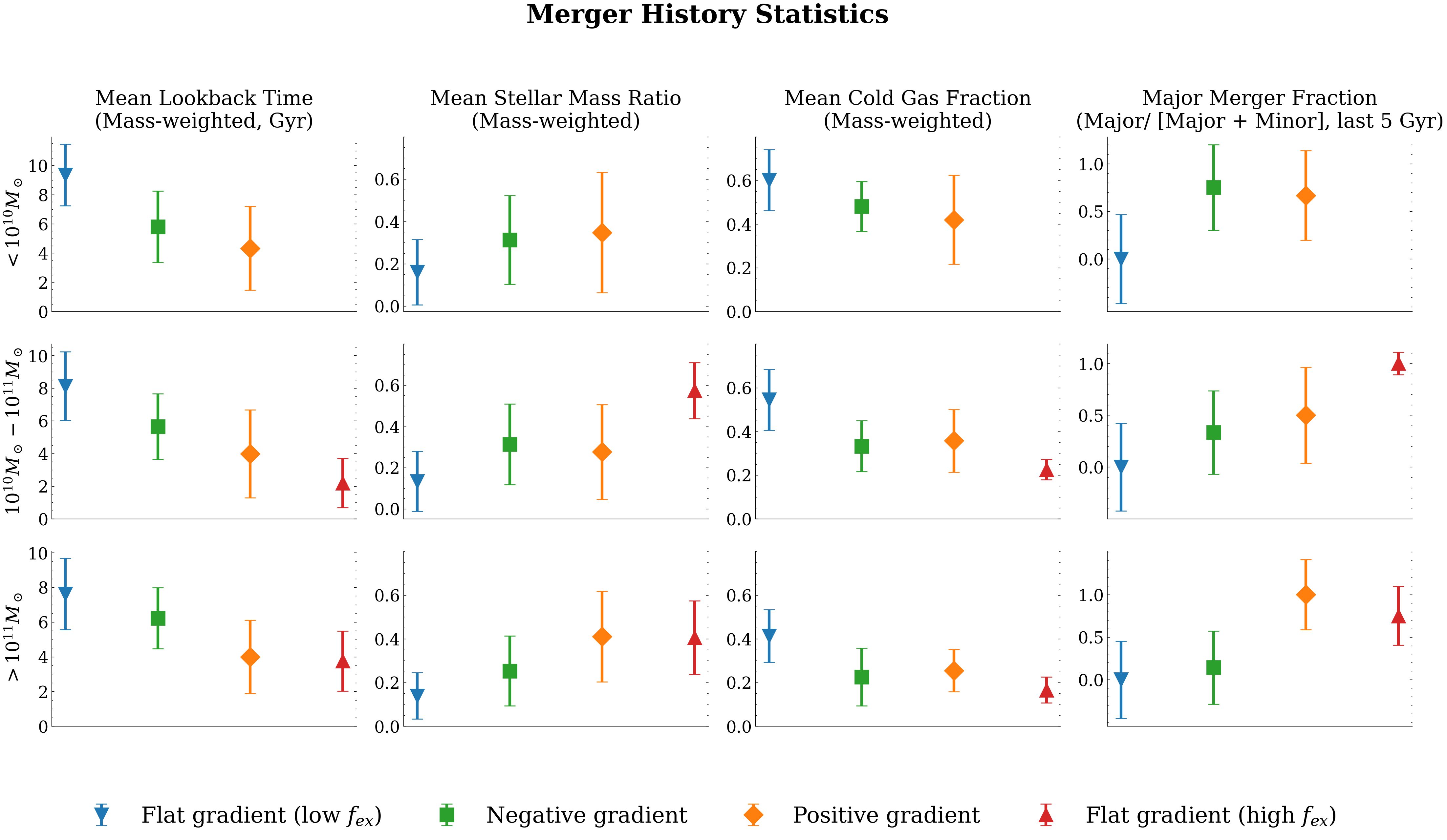}
    \caption{Distributions of key merger history metrics for MaNGIA galaxies, separated by ex-situ slope class. Top: Mean lookback time of mergers,
    weighted by the stellar mass of the secondary progenitors, reflecting when most mergers occurred. Second row: Mean cold gas fraction of merging companions, weighted by their stellar mass, indicating the typical gas richness of the merger history. Third row: Mean stellar mass ratio of mergers, also weighted by secondary progenitor mass, characterizing the relative strength of merger events. Bottom: Fraction of major mergers, calculated as the number of major mergers divided by the total number of major and minor mergers over the last 2 Gyr, highlighting the dominance of major versus minor mergers in recent assembly history.}
    \label{fig:merging_stats_mass_bins}
\end{figure*}

We recreate Fig. \ref{fig:merging_history_properties} from the main text, separating in stellar mass bins per category. We find that all trends still hold independent of the probed stellar mass, as shown in Fig. \ref{fig:merging_stats_mass_bins}.

\section{MaNGIA merging histories} \label{merger_histories_appendix}

\begin{figure}
    \centering
    \includegraphics[width=\linewidth]{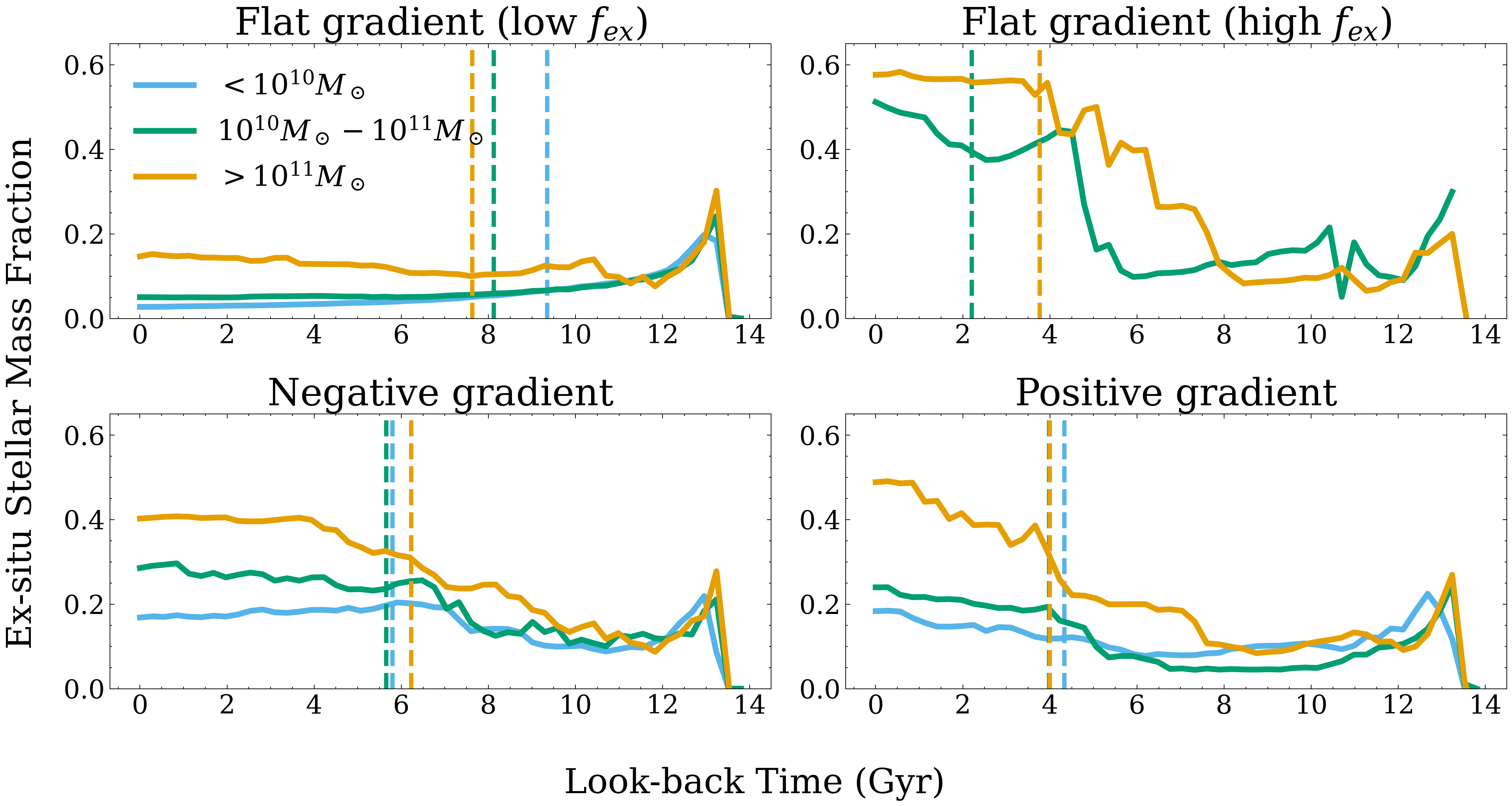}
    \caption{Stacked ex-situ stellar mass fraction histories for different slope classes in MaNGIA. Shown as a function of lookback time, grouped by present-day stellar mass. Each curve represents the median ex-situ fraction evolution for galaxies in a given class, showing distinct assembly histories associated with the inferred ex-situ radial gradient.}
    \label{fig:merging_history_groups}
\end{figure}

For each galaxy in MaNGIA, we attempt to reconstruct its merging history. For that, we follow its main progenitor branch across cosmic time, extracting both the ex-situ stellar mass and the total stellar mass at every available snapshot. We use the stellar mass assembly catalog by \citet{2016MNRAS.458.2371R}, which provide this information in a convenient format. From these, we compute the ex-situ stellar mass fraction at each redshift, thus reconstructing the time evolution of the relative contributions of accreted and in-situ stars for every galaxy. We then stack the individual histories per class and bin the results into three stellar mass groups, as shown in Figure \ref{fig:merging_history_groups}.

We find that the different classes exhibit distinct assembly histories, both in terms of merger timing and merger mass ratios. Flat gradient (low ex-situ) galaxies, dominated by in-situ star formation, show a remarkably quiet merger history across cosmic time, consistent with secular evolution and minimal contribution from accreted stars. Negative gradient galaxies tend to have experienced an early major merger (approximately 7–10 Gyr ago), after which their evolution is relatively quiescent. This early, significant accretion event likely built up a dense, ex-situ-dominated bulge, while subsequent star formation in the outskirts resulted in a younger, more in-situ-dominated disk.  Positive gradient galaxies, in contrast, display a more active and recent major merger history. This recent merging activity explains the high ex-situ stellar mass fraction in the outskirts, while gas inflows likely triggered central star formation, preserving the in-situ-dominated bulge. Finally, flat gradient (high ex-situ) galaxies appear to have undergone extremely major mergers throughout their history, resulting in a globally ex-situ-dominated stellar mass distribution across all radii.

\section{Star-formation histories}
\label{appendix_sfhs}
\begin{figure}
    \centering
    \includegraphics[width=\linewidth]{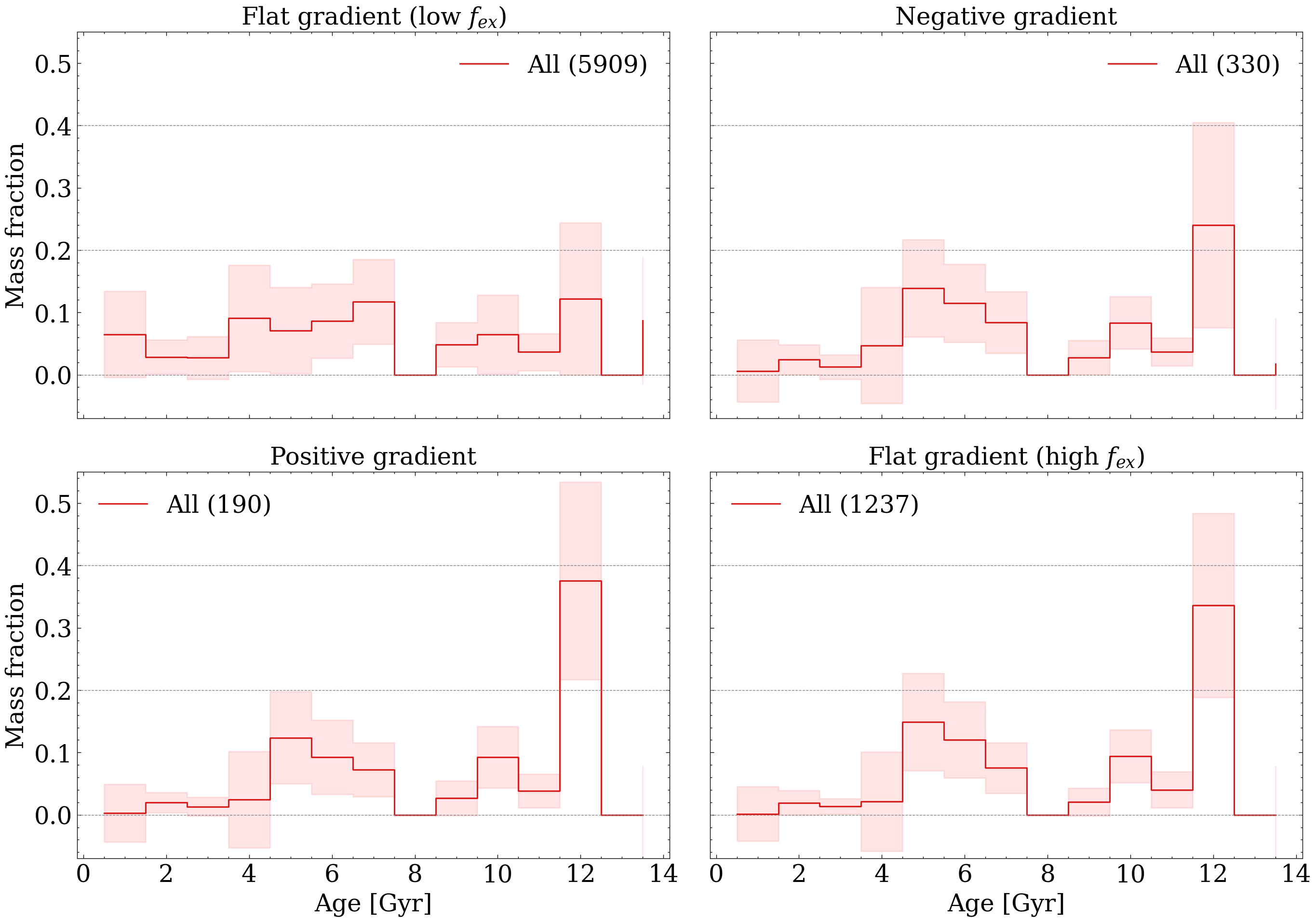}
   
    \caption{The star formation histories of the 4 different groups in MaNGA, as resolved by the Firely code. The solid line represents the median of the stacked star formation histories in each group, while the shaded regions cover 68\% percent of all data.}
    \label{fig:sfhs}
\end{figure}

We examine the star formation histories (SFHs) in the central regions ($r < 0.5R_e$) of MaNGA galaxies as resolved by the Firefly value-added catalog \citep{2022MNRAS.513.5988N, 2017MNRAS.466.4731G}. We stack the respective SFHs in each class based on the central ex-situ stellar mass fraction gradient to further assess differences in their assembly pathways and plot the results in Fig.~\ref{fig:sfhs}. While the differences in the stacked star-formation histories are subtle, there exist some trends consistent with the proposed scenarios. The flat gradient (low ex-situ) class tends to show continuous star formation, reflecting secular growth. The negative gradient class also shows a smooth history but exhibits two minor dual peaks in stellar age distributions, indicating an early burst followed by more recent star formation. Positive gradient and flat gradient (high ex-situ) classes display older stellar populations on average, with a stronger peak on early star formation episodes likely linked to merger-driven assembly. However, the differences between the two low ex-situ classes and between the two high ex-situ classes are subtle and their SFHs remain broadly consistent within 1-sigma uncertainties. Thus, while these trends support our evolutionary narrative, they should be regarded more as suggestive.

\end{appendix}

\end{document}